**Chapter 4: Modeling streaming potential in porous and fractured media, description and benefits of the effective excess charge density approach**

D. Jougnot (1), D. Roubinet (2), L. Guarracino (3), A. Maineult (1)

1. Sorbonne Université, CNRS, EPHE, UMR 7619 METIS, Paris, France

2. Geosciences Montpellier, UMR 5243, CNRS, University of Montpellier, France.

3. Facultad de Ciencias Astronomicas y Geofisicas, UNLP, CONICET, La Plata, Argentina

Corresponding autor: Damien Jougnot (*damien.jougnot@upmc.fr*)

**Abstract**

Self-potential signals can be generated by different sources and can be decomposed in various contributions. Streaming potential is the contribution due to the water flux in the subsurface and is of particular interest in hydrogeophysics and reservoir characterization. Being able to estimate water fluxes in porous and fractured media using streaming potential data relies on our understanding of the electrokinetic coupling at the mineral-solution interface and our capacity to understand, model, and upscale it. Two main approaches have been proposed to predict streaming potential generation in geological media. One of these approaches is based on determining the excess charge which is effectively dragged in the medium by water flow. In this chapter, we describe how to model the streaming potential by considering this effective excess charge density, how it can be defined, calculated and upscaled. We provide a short overview of the theoretical basis of this approach and we describe different applications to both water saturated and partially saturated soils and fractured media.

## 1. Introduction

Among geophysical methods, self-potential (SP) is considered to be one of the oldest as it can be tracked down to Robert Fox's work in 1830 (Fox, 1830). It consists in the passive measurement of the naturally occurring electrical field in the near surface. The minimum set-up to measure SP signals consists in two non-polarizable electrodes and a high impedance voltmeter. One of the electrodes is used as a reference while the other one is a rover electrode. The SP signal is to the electrical potential difference between those electrodes.

The SP method is relatively easy to set-up and data can be gathered quickly and easily. However, the extraction of useful information is a non-trivial task since the recorded signals are a superposition of different SP components. As S. Hubbard wisely wrote: "*Although self-potential data are easy to acquire and often provide good qualitative information about subsurface flows and other processes, a quantitative interpretation is often complicated by the myriad of mechanisms that contribute to the signal.*" (S. Hubbard in the foreword of Revil and Jardani, 2013). In natural porous media, SP signals are generated by charge separation that can have electrokinetic or electrochemical origins. In this chapter we only focus on the electrokinetic contribution to the SP signal: the streaming potential. More details regarding the SP method and its potential sources are described in details in Revil and Jardani (2013).

The electrokinetic (EK) contribution to the SP signal is generated from the water flow in porous media and the associated coupling with the mineral-solution interface. The surfaces of the minerals that constitute most geological media are generally electrically charged, which induce the development of an electrical double layer (EDL). This EDL contains an excess charge that counterbalances the charge of the mineral surfaces (see Hunter, 1981; Leroy and Revil, 2004). The EDL is generally composed of a Stern layer coating the mineral with a very limited thickness that only contains counterions (i.e., ions with an opposite electrical charge compare to the surface charges) and a diffuse layer that contains both counterions and co-ions but with a net excess charge (Fig. 1a). We call shear plane the separation between the mobile and immobile parts of the water molecules when subjected to a pressure gradient and we call ζ-potential the electrical potential characterizing it (see Hunter, 1981). It is often approximated as the limit between the Stern and diffuse layers (e.g., Leroy and Revil, 2004). When, submitted to a pressure gradient, the water flows in the pore space, dragging a fraction of the excess charge. This phenomenon gives rise to the socalled streaming current and a resulting electrical potential field (i.e., the streaming potential).

The first experimental descriptions of the streaming potential can be found in Quincke (1861) and later Dorn (1880). Helmholtz (1879) and von Smoluchowski (1903) proposed a theoretical description of the electrokinetic phenomena by considering a water-saturated capillary and by defining the streaming potential coupling coefficient as the ratio between the pressure and the electrical potential differences at the boundaries of the capillary. The so-called Helmholtz-Smoluchowski (HS) equation relates this coupling coefficient to the properties of the pore solution. This equation is independant from the medium geometrical properties and has therefore been used for any kind of medium. It is valid as long as the electrical conductivity of the mineral surface can be neglected. Alternative equations have been proposed by several researchers when this assumption cannot be made (e.g., Revil et al., 1999; Glover and Déry, 2010). The use of the Helmholtz-Smoluchovski (HS) equation to determine the streaming potential coupling coefficient has been proven very useful for a wide range of materials fully saturated with water (e.g., Pengra et al. 1999, Jouniaux and Pozzi, 1995). However, the HS equation cannot be applied for partially saturated conditions and the evolution of the streaming potential coupling coefficient when the water saturation decreases is still the subject of important debates in the community (e.g., Allègre et al. 2014, Fiorentino et al. 2016, Zhang et al. 2017).

An alternative approach to model the electrokinetic coupling phenomena is based on the excess charge located in the EDL which is dragged by the water flow in the pore space. It was first formulated by Kormiltsev et al. (1998) as the electrokinetic coefficient, and later physically developed by Revil and co-workers using different up-scaling methods (e.g., Revil and Leroy, 2004; Linde et al. 2007; Revil et al. 2007; Jougnot et al. 2012). This chapter aims at describing the theory and the usefulness of the effective excess charge density approach to better understand and model the generation of the streaming potential. First, the theory of this approach will be described, linking it to the more traditional approach that uses the coupling coefficient. Then, the evolution of the effective excess charge with different rock properties and environmental variables will be studied. Finally, this approach will be used to simulate the generation of the streaming potential in two complex media: a partially saturated soil and a fractured domain.

## 2. Theory

### *2.1. Description of the electrical double layer*

Figure 1a is a schematic description of the EDL that develops at the interface between a charged mineral and the pore water solution. The amount and the sign of the surface charge can vary from one mineral to another or with varying pH (e.g., Leroy and Revil, 2004). We here call $Q_0$ the surface charge of the mineral (in C m$^{-2}$) that are counterbalanced by the charge (i.e., counterions) located in the EDL. These counterions are distributed between: (1) the Stern layer, sometimes called fixed layer as the ions are sorbed onto the mineral surface, and (2) the diffuse layer (also called Gouy–Chapman layer), where ions are less affected by the surface charges and can diffuse more freely. At thermodynamic equilibrium and in saturated conditions, these charges respect the following charge balance equation:

$$\frac{S_{sw}}{V_w}\left(Q_0 + Q_\beta\right) + \bar{Q}_v = 0, \tag{1}$$

where $S_{sw}$ is the surface of the mineral (in m$^2$), $V_w$ is the water volume in the pore space (in m$^3$), $Q_\beta$ is the charge of the Stern layer (in C m$^{-2}$), and $\bar{Q}_v$ is the volumetric charge density in the diffuse layer (in C m$^{-3}$). In partially saturated conditions, that is, when the pore space contains air and water, an additional interface and electrical double layer are present in the porous media (e.g., Leroy et al. 2012). The specific surface area of the air-solution interface is considered to be negligible by many authors compared to the mineral-solution one (e.g., Revil et al. 2007; Linde et al. 2007). However, some works have recently challenged this hypothesis (e.g., Allègre et al. 2015; Fiorentino et al. 2017).

While the Stern layer contains only counterions and has negligible thickness, the diffuse layer contains both counterions and co-ions and its thickness strongly depends on the pore solution chemistry. The distribution of ions in the diffuse layer is determined by the local electrical potential $\psi$ distribution as a function of the distance from the shear plane, $x$:

$$\psi(x) = \zeta \exp\left(-\frac{x}{l_D}\right), \tag{2}$$

where $\zeta$ is the so-called zeta potential (in V), the local electrical potential at the shear plane, and $l_D$ is the Debye length (in m) defined as:

$$l_D = \sqrt{\frac{\varepsilon_w k_B T}{2 N_A I e_0^2}}, \tag{3}$$

where $\varepsilon_w$ is the dielectric permittivity of the pore water (in F m$^{-1}$), $k_B = 1.381\times10^{-23}$ J K$^{-1}$ is the Boltzmann constant, $T$ is the temperature (in K), $N_A$ is the Avogadro number (in mol$^{-1}$), $I$ is the ionic strength of the pore water solution (in mol L$^{-1}$), and $e_0 = 1.6\times10^{-19}$ C is the elementary charge. The ionic strength of an electrolyte is given by

$$I = \frac{1}{2}\sum_{i=1}^{N} z_i^2 C_i^0 , \tag{4}$$

where $N$ is the number of ionic species $i$, $z_i$ and $C_i^0$ are the valence and the concentration (in mol $L^{-1}$) of the $i^{th}$ ionic species. More precisely, $C_i^0$ is the concentration of the ionic species outside the EDL (i.e., in the free electrolyte). In the diffuse layer, and under the assumption that the pores are larger than the diffuse layer (i.e., thin layer assumption), the concentration of each ionic species follows:

$$C_i(x) = C_i^0 \exp\left(-\frac{z_i e_0 \psi(x)}{k_B T}\right). \tag{5}$$

The excess charge distribution in the diffuse layer can be expressed by the sum of charges from each species (see Fig. 1b):

$$\bar{Q}_v(x) = N_A \sum_{i=1}^{N} z_i e_0 C_i(x). \tag{6}$$

From the above equations, it becomes easy to see that the thickness of the diffuse layer is related to the Debye length. The diffuse layer extension corresponds to the fraction of the pore space for which a significant amount of excess charge is not negligible: i.e., roughly 4 $l_D$ (Fig. 1b).

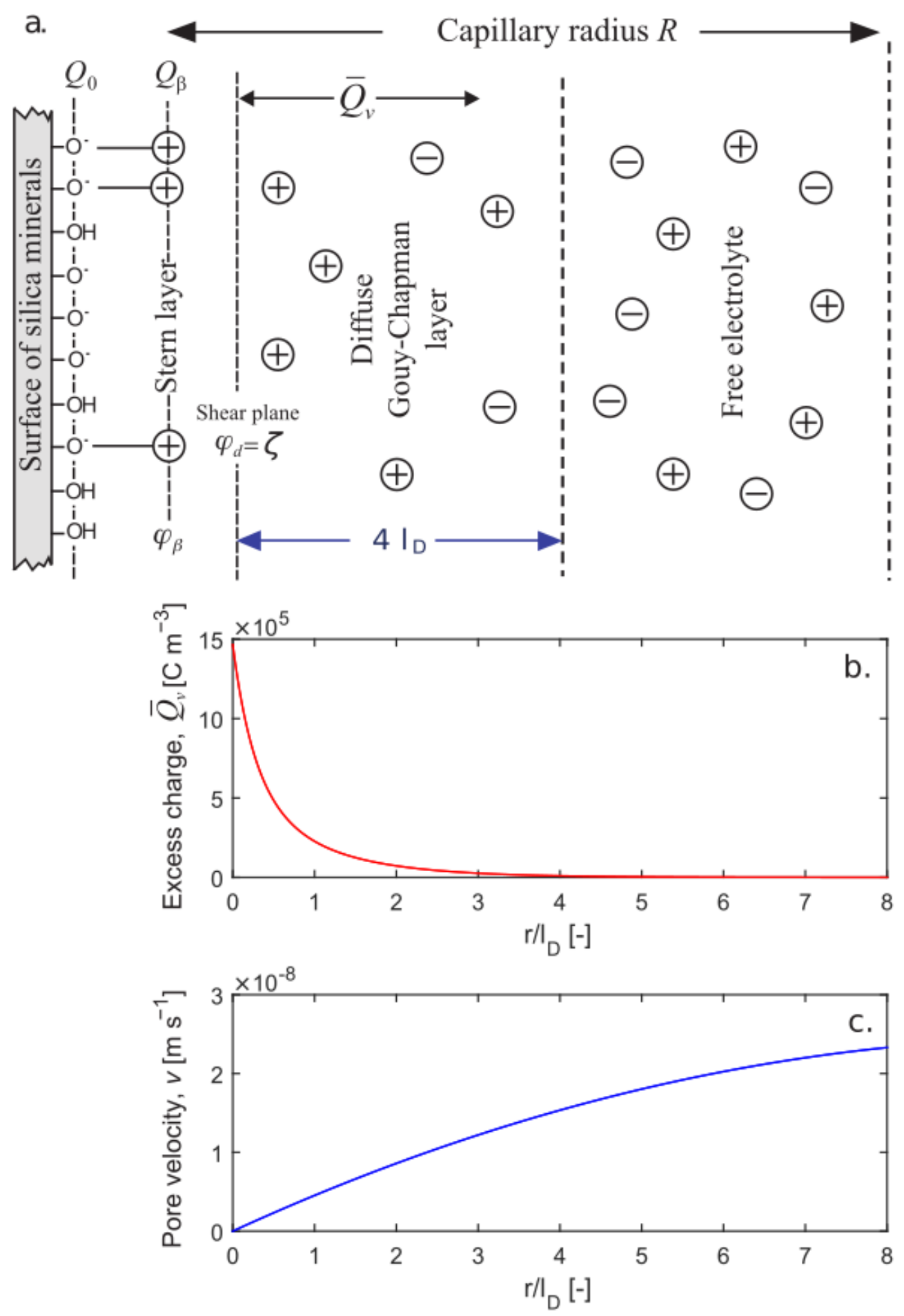

*Figure 1: (a) Sketch of the electrical double layer. Distribution of (b) the excess charge and (c) the pore water velocity as a function of the distance from the shear plan (modified from Jougnot et al. 2012).*

### *2.2. Electrokinetic coupling framework*

The constitutive equations describing the coupling between the electrical field and the water flow can be written as follow (e.g., Nourbehecht, 1963):

$$\begin{bmatrix} \mathbf{j} \\ \mathbf{u} \end{bmatrix} = -\mathbf{L} \begin{bmatrix} \nabla \varphi \\ \nabla \left( p_w - \rho_w g z \right) \end{bmatrix} \tag{7}$$

where $\mathbf{j}$ is the electrical current (in A m$^{-2}$), $\mathbf{u}$ is the water flux (in m s$^{-1}$), $\varphi$ is the electrical potential, $p_w$ is the water pressure (in Pa), $g$ is the gravitational constant (in m s$^{-2}$), z is the elevation (in m) and $\rho_w$ the water density (kg m$^{-3}$). The coupling matrix $\mathbf{L}$ is defined as:

$$\mathbf{L} = \begin{bmatrix} \sigma & L^{EK} \\ L^{EK} & \dfrac{k}{\eta_w} \end{bmatrix} \tag{8}$$

where $\sigma$ is the electrical conductivity of the medium (in S m$^{-1}$), $k$ is the medium permeability (in m$^2$), and $\eta_w$ is the dynamic viscosity of the water (Pa s). From this coupling matrix, one can easily identify the Ohm's law and the Darcy's law through $L_{11}$ (i.e., $\sigma$) and $L_{22}$ (i.e., $k/\eta_w$), respectively. Following Onsager (1931), the two non-diagonal terms should be equal and correspond to the electrokinetic coupling coefficient $L^{\mathrm{EK}}$. It can be used to describe both the electrokinetic coupling (i.e., a water flow induces an electrical current) and the electro-osmotic coupling (i.e., an electrical current induces a water flow) in porous media. However, for most environmental applications (except for compacted clay rocks), the effect of electro-osmosis on the water flow can be safely neglected (e.g., Revil et al., 1999). In this case, the system can be simplified by neglecting $L_{21}$:

$$\mathbf{j} = -\sigma\nabla\varphi - L^{EK}\nabla\left(p_w - \rho_w g z\right), \tag{9}$$

$$\mathbf{u} = -\frac{k}{\eta_w}\nabla\left(p_w - \rho_w g z\right). \tag{10}$$

Using this simplification and considering that there is no external current in the system (i.e., no current injection, and thus $\nabla\cdot\mathbf{j} = 0$), Sill (1983) proposes the following Poisson's equation for describing the streaming potential generation:

$$\nabla\cdot\left(\sigma\nabla\varphi\right) = \nabla\cdot\mathbf{j}_S, \tag{11}$$

where $\mathbf{j}_S$ is the streaming current density (in A m$^{-2}$) resulting from the electrokinetic coupling phenomenon that can be written as:

$$\mathbf{j}_S = -L^{EK}\nabla\left(p_w - \rho_w g z\right). \tag{12}$$

Note that Eq. (12) is often expressed as a function of the hydraulic head gradient H (in m), which yields to:

$$\mathbf{j}_S = -L^{EK}\rho_w g\nabla H, \tag{13}$$

with $H = \dfrac{p_w}{\rho_w g} + z$ (in m).

Based on the simple geometry of a capillary tube, Helmholtz (1879) and von Smoluchowski (1903) developed a simple equation to quantify the electrokinetic coupling coefficient $L^{\mathrm{EK}}$ and proposed the Helmholtz-Smoluchowski (HS) equation, defining the coupling coefficient $C^{\mathrm{HS}}$ (in V Pa$^{-1}$):

$$C^{HS} = \frac{L^{EK}}{\sigma} = \frac{\varepsilon_w\zeta}{\eta_w\sigma_w}, \tag{14}$$

where $\sigma_w$ is the pore water electrical conductivity (in S m$^{-1}$). See also the complete derivation in Rice and Whitehead (1965). The HS equation has proven to be very useful as, in absence of external current, it relates the electrical potential difference $\Delta\varphi$ that can be measured at the boundaries of a sample to the pressure difference $\Delta p_w$ to which it is submitted:

$$C^{HS} = \left.\frac{\Delta\varphi}{\Delta p_w}\right|_{\mathbf{j}=0}. \tag{15}$$

However, Eq. (14) is only valid when the surface conductivity of the minerals can be neglected. When it is not the case, modified versions of Eq. (14) have been proposed in the literature (e.g., Revil et al. 1999; Glover and Déry, 2010). Another limitation with the HS coupling coefficient is to consider a porous medium under partially saturated conditions. Many models have been proposed to describe the evolution of the coupling coefficient with variable water saturation (e.g., Perrier and Morat 2000, Guichet et al. 2003, Revil and Cerepi 2004, Allegre et al. 2010, 2015). Nevertheless, as illustrated in Zhang et al. (2017) (their Fig. 1), no consensus has been found on the behavior of the coupling coefficient as a function of water saturation as it seems to differ from one medium to another.

In order to deal with these two issues (i.e., surface conductivity and partially saturated media), an alternative approach can be used to describe the coupling coefficient. In this case, the electrokinetic coupling variable becomes the excess charge which is effectively dragged by the water flow in the pore space.

### *2.3. From the coupling coefficient to excess charge*

Kormiltsev et al. (1998) is the first English reference proposing to re-write Eq. (12) using a different coupling variable. In their new formulation, they relate the source current density directly to the average water velocity in the porous medium. Indeed, combining the definition of the Darcy velocity (Eq. 10) and the electrokinetic source current density (Eq. 12), it is possible to propose a variable change such as:

$$\mathbf{j}_s = L^{EK}\frac{\eta_w}{k}\mathbf{u}, \tag{16}$$

where the middle term $L^{EK}\frac{\eta_w}{k}$ is expressed in C m$^{-3}$ and corresponds to a volumetric excess charge as defined in section 2.1. It is therefore possible to re-write Eq. (12) as:

$$\mathbf{j}_S = \hat{Q}_v\mathbf{u}, \tag{17}$$

where $\hat{Q}_v$ (in C m$^{-3}$) is the volumetric excess charge which is effectively dragged by the water flow in the pore space (called $\alpha$ in Kormiltsev et al., 1998). Independently from Kormiltsev et al. (1998), Revil and Leroy (2004) developed a theoretical framework for various coupling properties based on this effective excess charge approach for saturated porous media. In this work, a formulation for the electrokinetic coupling coefficient is given as an alternative to the HS coupling coefficient (Eq. 14):

$$C^{EK} = -\frac{\hat{Q}_v k}{\sigma \eta_w}. \tag{18}$$

This formulation is of interest to relate the coupling coefficient to the permeability and the electrical conductivity of the medium, two parameters that can be measured independently. Later, Revil et al. (2007) and Linde et al. (2007) extended this framework to describe the electrokinetic coupling in partially saturated media, considering that the different parameters on which depends the coupling coefficient are function of the water saturation, $S_w$,

$$C^{EK}(S_w) = -\frac{\hat{Q}_v(S_w) k^{rel}(S_w) k}{\sigma(S_w)\eta_w}, \tag{19}$$

with $k^{rel}(S_w)$ the relative permeability function comprised between 0 and 1. In the following, the upper script *rel* refers to the value of a parameter relatively to its value under fully water saturated conditions.

Following the definition of Guichet et al. (2003), the relative coupling coefficient $C_{rel}^{EK}$ (unitless) can then be expressed as relative to the value in saturated conditions ($C_{sat}^{EK}$) which yield to (Linde et al., 2007; Jackson, 2010):

$$C_{rel}^{EK}(S_w) = \frac{C^{EK}(S_w)}{C_{sat}^{EK}} = \frac{\hat{Q}_v^{rel}(S_w) k^{rel}(S_w)}{\sigma^{rel}(S_w)}. \tag{20}$$

From Eq. (19), it is interesting to note that the coupling coefficient results from the product of three different petrophysical properties of the porous medium: $k$, $\sigma$, and $\hat{Q}_v$. Therefore, the coupling coefficient strongly depends on these parameters and their evolution. The permeability, $k$, and the electrical conductivity, $\sigma$, are two extensively studied properties that have been shown to vary by orders of magnitude between the different lithologies, but also for varying water saturation and, for $\sigma$, different pore water conductivities.

Various petrophysical relationships exist to describe $k$ and $\sigma$. The permeability can be expressed as a function of the porosity and the medium tortuosity (e.g., Kozeny, 1927; Carman, 1937; Soldi et al., 2017) or the water saturation (e.g., Brooks and Corey 1964, van Genuchten 1980, Soldi et al., 2017). On the other side, the electrical conductivity depends on the porosity, the water saturation and the pore water conductivity (e.g., Archie, 1942; Waxman and Smits, 1984; Linde et al., 2006). However, the evolution of the effective excess charge density still remains unknown. The present contribution aims at better describing this property, its evolution, and its usefulness to understand and model the streaming current generation in porous and fractured media.

## *2.4. Determination of the effective excess charge density*

*2.4.1 Under water saturated conditions*

The determination of the effective excess charge density has been the subject of only a couple of studies during the last two decades. One can identify two main ways to determine this crucial parameter: (1) empirically from experimental measurements and (2) numerically or analytically through an up-scaling procedure.

Based on previous studies from the literature and the theoretical framework described by Kormiltsev et al. (1998), Titov et al. (2001) first showed that $\hat{Q}_v$ strongly depends on the medium permeability. Then, Jardani et al. (2007) proposed a very useful and effective empirical relationship:

$$\log(\hat{Q}_v) = A_1 + A_2 \log(k), \tag{21}$$

where $A_1$ = -9.21 and $A_2$ = -8.73 are constant values obtained by fitting Eq. (21) to a large set of experimental data. This relationship has been shown to provide a fairly good first approximation for all kinds of water saturated porous media that range from gravels to clay (Fig. 2). Note that other empirical relationships can be found in the literature (e.g., Bolève et al. 2012). Linking $\hat{Q}_v$ to the permeability seems fairly logical since both properties depend on the interface between mineral and solution: the permeability through viscous energy dissipation and the effective excess charge density through the EDL. However, the use of this relationship is limited by the fact that it does not take into account other physical properties like porosity and the chemical composition of the pore water. This particular point has been discussed by Jougnot et al. (2015) while modeling the SP response of a saline tracer infiltration in the near surface.

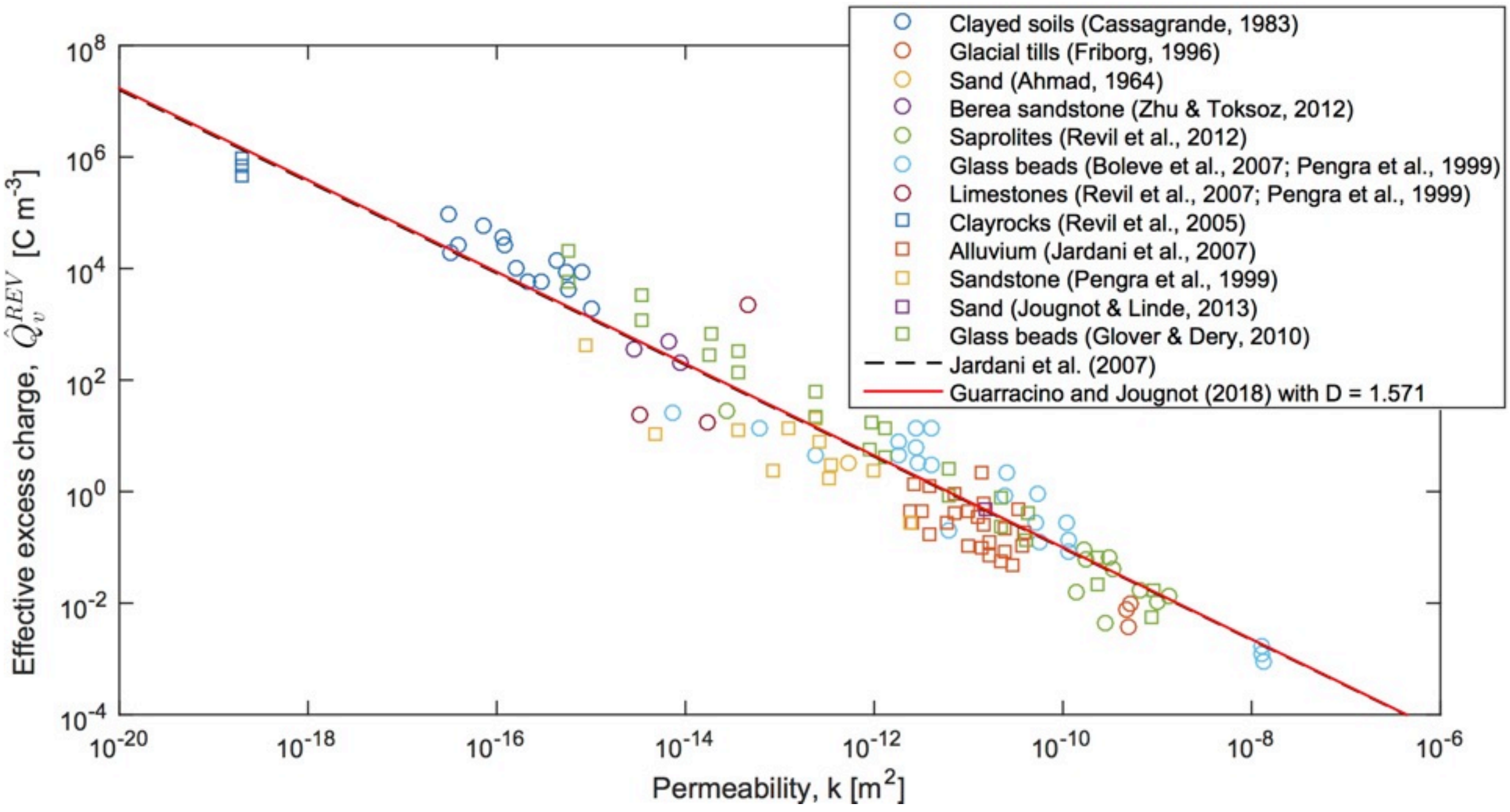

*Figure 2: Effective excess charge density of various porous media as a function of the permeability (modified from Guarracino and Jougnot, 2018).*

The second approach to obtain the effective excess charge density is through an up-scaling procedure. In this approach the transport of the excess charge density by the water flux in the medium is explicitly considered. In order to perform this up-scaling, one must simplify the problem using geometrical approximations to describe the porous medium. Following the original work of von Smoluchovski (1903), it is possible to consider the electrokinetic coupling phenomena occurring in a capillary (e.g., Rice and Whitehead, 1951; Packard, 1953) or in a bundle of capillaries (e.g., Bernabé, 1998; Jackson, 2008, 2010; Jackson and Leinov, 2012).

More recently, Guarracino and Jougnot (2018) proposed an analytical mechanistic model to determine the effective excess charge under saturated conditions for a bundle of capillaries. This model is based on a two-steps up-scaling procedure that was proposed numerically by Jougnot et al. (2012): (1) from the EDL scale to the effective excess charge in a single capillary and then (2) from one capillary to a bundle of capillaries (i.e. the REV).

Based on the EDL description and assumptions presented in section 2.1, Guarracino and Jougnot (2018) derived a closed-form equation for the effective excess charge density in a single capillary with the radius $R$ (in m), $\hat{Q}_v^{\mathrm{R}}$ (in C m$^{-3}$):

$$\hat{Q}_v^{\mathrm{R}}(R) = \frac{8 N_{\mathrm{A}} e_0 C_{\mathrm{w}}^0}{\left(R/l_{\mathrm{D}}\right)} \left[ -2\frac{e_0 \zeta}{k_{\mathrm{B}} T} - \left( \frac{e_0 \zeta}{3 k_{\mathrm{B}} T} \right)^3 \right]. \tag{22}$$

Then, by considering a fractal law for the pore size distribution, that is a power law distribution relating the pore size $R$ to the number of pores in the medium $N(R)$ (e.g., Guarracino et al., 2014; Tyler & Wheatcraft, 1990; Yu et al., 2003):

$$N(R) = \left( \frac{R^{\mathrm{REV}}}{R} \right)^D, \tag{23}$$

where $D$ is the fractal dimension (unitless), they derived a closed-form equation to determine the effective excess charge density at the scale of the representative elementary volume (REV) (i.e., the bundle of capillaries), $\hat{Q}_v^{\mathrm{REV}}$ (in C m$^{-3}$):

$$\hat{Q}_v^{\mathrm{REV}} = N_{\mathrm{A}} e_0 C_{\mathrm{w}}^0 l_{\mathrm{D}} \left[ -2\frac{e_0 \zeta}{k_{\mathrm{B}} T} - \left( \frac{e_0 \zeta}{3 k_{\mathrm{B}} T} \right)^3 \right] \frac{1}{\tau^2} \frac{\phi}{k}, \tag{24}$$

where the parameters controlling $\hat{Q}_v^{\mathrm{REV}}$ can be decomposed in two main parts (1) the geometrical properties (i.e., petrophysical properties): porosity $\phi$, permeability $k$, and hydraulic tortuosity $\tau$ and (2) the electro-chemical properties: ionic concentration, Debye length, and Zeta potential. Note that all these properties can be estimated independently. By arranging Eq. (24), it is possible to derive the empirical relationship proposed by Jardani et al. (2007) (Eq. 21) and to obtain expression for the fitting constants $A_1$ and $A_2$ in terms of fractal dimension and chemical parameters. The performance of the model is tested with the extensive data set presented in Fig. 2.

*2.4.2 Under partially saturated conditions*

Under partially saturated conditions, that is, when the water volume in the pore space diminishes, the behavior of the effective excess charge is still under discussion. One could see two different up-scaling approaches to determine it: (1) the volume averaging approach and (2) the flux-averaging approach.

The volume averaging approach to determine the evolution of $\hat{Q}_v^{\mathrm{REV}}(S_{\mathrm{w}})$ was first proposed by Linde et al. (2007) to explain the data from a sand column drainage experiment and described in detail by Revil et al. (2007) in a very complete electrokinetic framework in partially saturated porous media. This up-scaling approach is built on the fact that no matter the medium water saturation, the surface charge to counterbalance is constant. That is, when the water volume decreases, the total excess charge diminishes but its density increases linearly. It yields:

$$\hat{Q}_v^{\mathrm{REV}}(S_{\mathrm{w}}) = \frac{\hat{Q}_v^{\mathrm{REV,sat}}}{S_{\mathrm{w}}}. \tag{25}$$

This approach has been successfully tested experimentally in various works mainly on sandy materials (e.g., Linde et al. 2007, Mboh et al. 2012, Jougnot and Linde 2013). However, when applied to more complex soils, Eq. (25) seems to fail reproducing the magnitudes observed.

Considering the porous medium as a bundle of capillaries provides a theoretical tool to perform the up-scaling of electrokinetic properties under partial saturation. Jackson (2008, 2010) and Linde (2009) propose different models to determine the evolution of the coupling coefficient with varying water saturation. The distribution of capillary sizes in the considered bundle is a way to take the heterogeneity of the pore space into account in the models. Building on the previous works cited above, Jougnot et al. (2012) propose a new way to numerically determine the evolution of the effective excess charge as a function of saturation. The numerical up-scaling proposed by these authors is called flux averaging approach, by opposition to the volume averaging one (Eq. 25), as it is based on the actual distribution of the water flux in the pore space and therefore on the fraction of the excess charge that is effectively dragged by it. The model can be expressed by:

$$\hat{Q}_v^{\mathrm{REV}}(S_{\mathrm{w}}) = \frac{\int_{R_{\min}}^{R_{\mathrm{Sw}}} \hat{Q}_v^{\mathrm{R}}(R) v^R(R) f_D(R) dR}{\int_{R_{\min}}^{R_{\mathrm{Sw}}} v^R(R) f_D(R) dR}, \tag{26}$$

where $\hat{Q}_v^{\mathrm{R}}(R)$ is the effective excess charge density (in C m$^{-3}$) in a given capillary $R$ as expressed by Eq. (21), $v^R(R)$ is the pore water velocity in the capillary (in m s$^{-1}$), and $f_D(R)$ is the capillary size distribution of the considered medium. Although this flux-averaging model can consider any kind of capillary size distribution, Jougnot et al. (2012) propose to infer $f_D(R)$ from the hydrodynamic properties of the considered porous medium. It yields two approaches: (1) the water retention (WR) and (2) the relative permeability (RP) based on the corresponding hydrodynamic functions. From various studies, it has been shown that the WR approach tends to better predict the relative evolution of the effective excess charge density as a function of saturation, while the RP approach performs better for amplitude prediction (e.g., Jougnot et al. 2012, 2015). Therefore, following the proposition of Jougnot et al. (2015), we suggest that the effective excess charge density under partially saturated conditions can be obtained by:

$$\hat{Q}_v^{\mathrm{REV}}(S_{\mathrm{w}}) = \hat{Q}_v^{\mathrm{REV,rel}}(S_{\mathrm{w}})\hat{Q}_v^{\mathrm{REV,sat}}, \tag{27}$$

where the saturated effective excess charge density $\hat{Q}_v^{\mathrm{REV,sat}}$ can be obtained from Eq. (24) and the relative excess charge density $\hat{Q}_v^{\mathrm{REV,\,rel}}(S_{\mathrm{w}})$ can be determined using Eq. (26).

It is worth noting that Jougnot and Linde (2013) shown that the predictions of Eq. (25) and (26) can overlap over a large range of saturation for certain sandy materials (e.g., the one used in Linde et al. 2007), which explains why the volume averaging model performed well in Linde et al. (2007) and possibly in Mboh et al. (2012) as they used a similar material.

## 3. Evolution of the effective excess charge

### *3.1 Evolution with the salinity*

From the theory section, it clearly appears that the pore water salinity strongly influences the electrokinetic coupling. Indeed, the pore water electrical conductivity explicitly appears in the coupling coefficient definition (Eqs. 14 and 18). Nevertheless, the pore water salinity also strongly affects the properties of the EDL. Eq. (3) shows its effect on the extension of the diffuse layer, while many studies show that it also changes the value of the $\zeta$-potential (e.g., Pride and Morgan, 1991; Jaafar et al., 2009; Li et al., 2016). In the present approach, we use the Pride and Morgan (1991) model:

$$\zeta(C_w^0) = a + b\log(C_w^0), \tag{28}$$

where a = −6.43 mV and b = 20.85 mV for silicate-based materials and for NaCl brine according to Jaafar et al. (2009) if $\zeta$ is expressed in mV and $C_w^0$ in mol $L^{-1}$. Note that the behavior of the $\zeta$-potential as a function of the salinity is challenged in the literature (e.g., see the discussion in Fiorentino et al., 2016).

Figure 3 illustrates the evolution of the effective excess charge density as a function of the pore water salinity (i.e., ionic concentration of NaCl). The experimental data come from the study of Pengra et al. (1999) for different porous media, while the model is the one proposed by Guarracino and Jougnot (2018) where the hydraulic tortuosity (i.e., the only parameter not measured by Pengra et al., 1999) is optimized to fit the data. The overall fit is pretty good, indicating that the Guarracino and Jougnot (2018) model correctly takes into account the effect of the salinity on the EDL and resulting effective excess charge density.

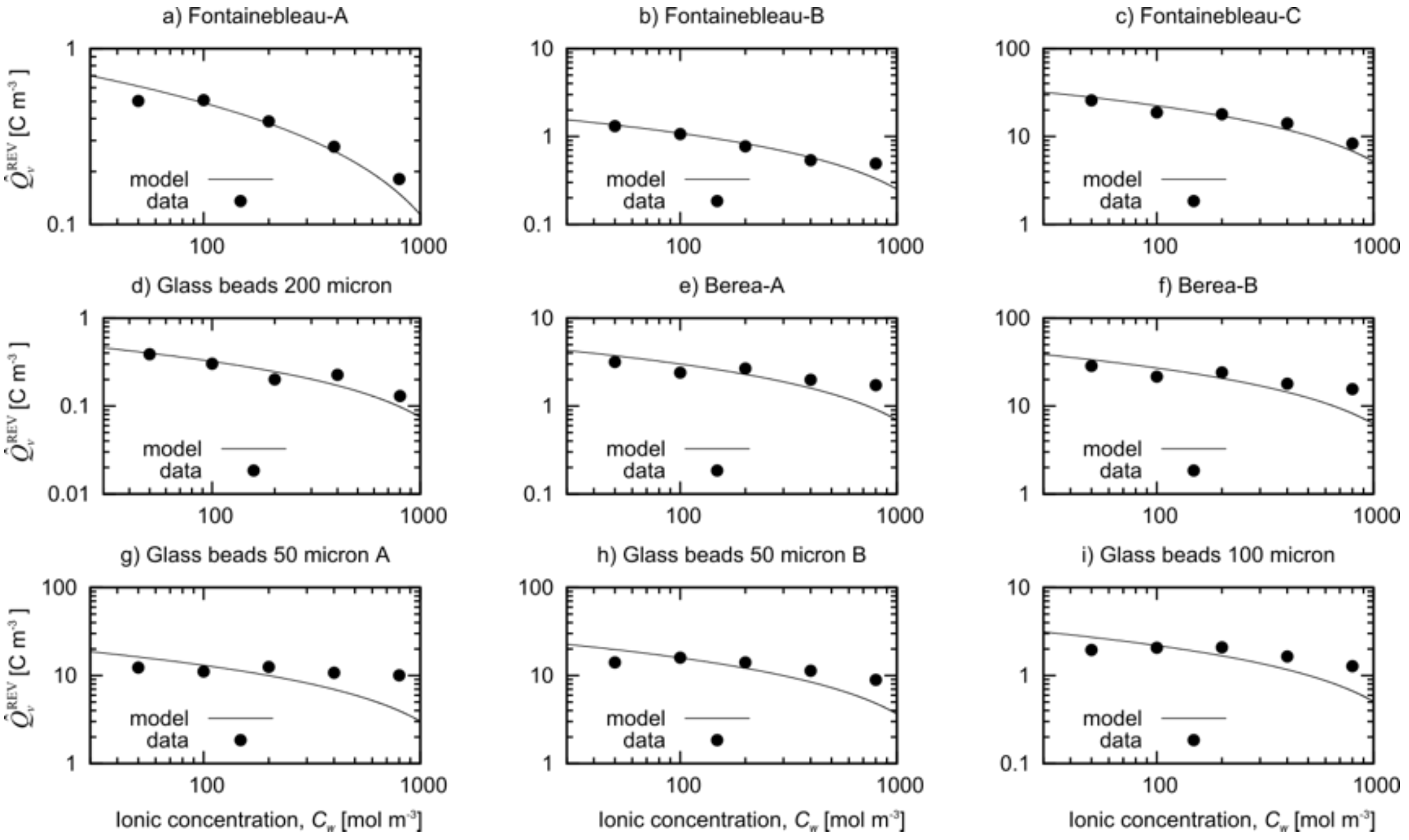

*Figure 3: Effective excess charge density of various porous media as a function of the ionic concentration of the NaCl in the pore water. The experimental data have been extracted from Pengra et al. (1999).*

### *3.2 Evolution with the petrophysical properties*

From previous section, it is clear that the effective excess charge density is dependent on petrophysical properties like permeability, porosity, and hydraulic tortuosity. In contrast to other models, Guarracino and Jougnot (2018) explicitly express $\hat{Q}_v^{\mathrm{REV}}$ as a function of these three parameters.

Glover and Déry (2010) conducted a series of electrokinetic coupling measurements on well-sorted glass bead samples of different radii at two pore water salinities. They also performed an extensive petrophysical characterization of each sample, providing all the necessary parameters to test the model proposed by Guarracino and Jougnot (2018), except for the hydraulic tortuosity. Figure 4a shows the $\hat{Q}_v^{\mathrm{REV}}$ predicted by this model (using $\tau$ = 1.2) and by the empirical relationship from Jardani et al. (2007) (Eq. 21). Figure 4b compares the coupling coefficient measured by Glover and Déry (2010) with the coupling coefficients calculated using the $\hat{Q}_v^{\mathrm{REV}}$ predicted by the models of Guarracino and Jougnot (2018) and Jardani et al. (2007), respectively. One can see that the model informed by the measured petrophysical parameter performs better and is able to reproduce the entire dataset with a single value of hydraulic tortuosity. A better fit can be obtained by optimizing the hydraulic tortuosity for each sample.

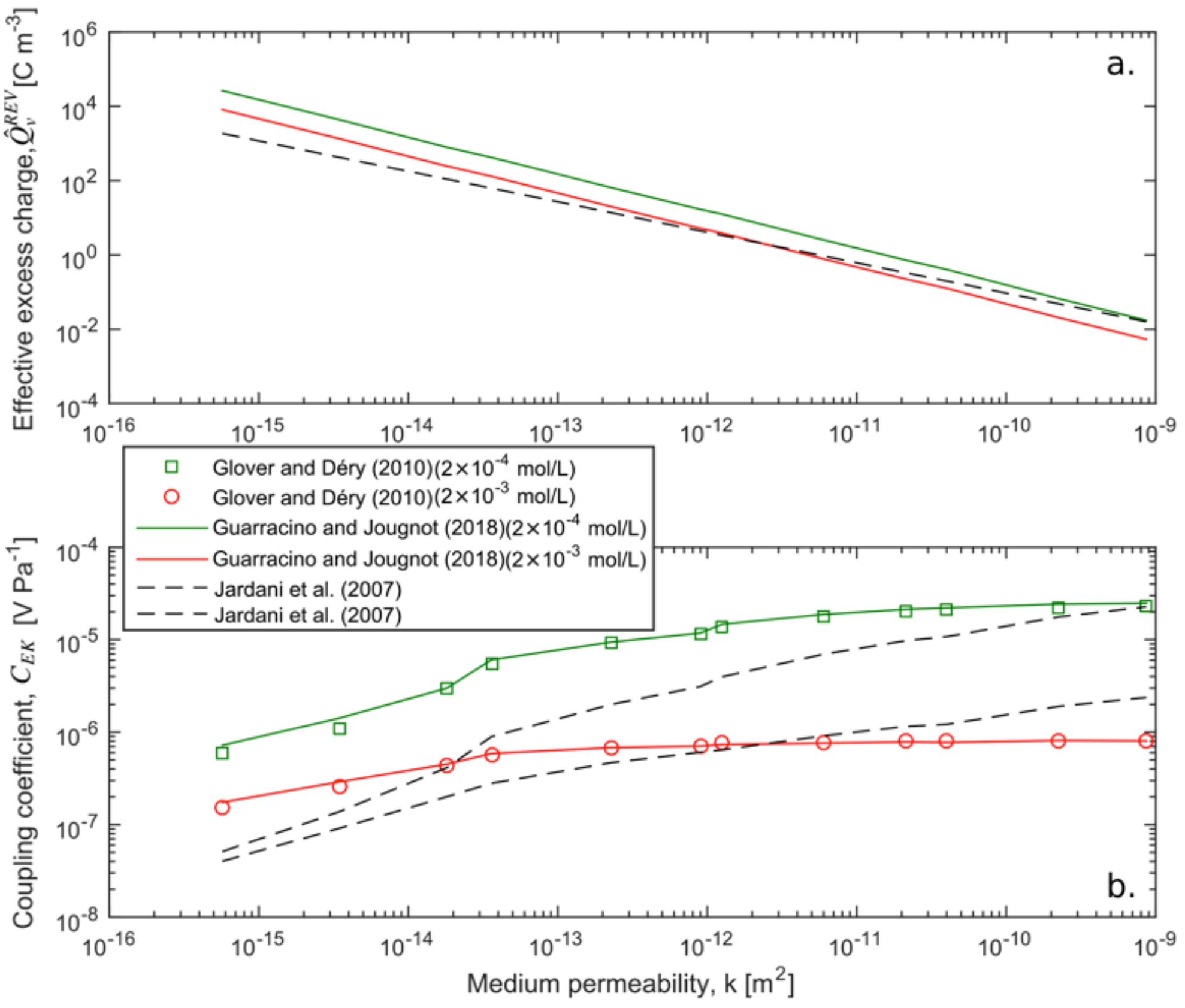

*Figure 4: Effective excess charge density of various porous media as a function of the permeability for $\tau$ = 1.2.*

The link between effective excess charge density and hydraulic tortuosity can be explicitly seen in Eq. (24). Unfortunately, the hydraulic tortuosity is not an easy parameter to measure for all type of porous media; Clennell (1997) provides an extensive review of the different definitions and models to estimate tortuosities in porous media. Among others, Windsauer et al. (1952) proposes a simple way to relate the hydraulic tortuosity to the formation factor $F$, which is easier to measure:

$$\tau_e = \sqrt{F\phi}\,, \tag{29}$$

where $\phi$ is the porosity of the medium.

Figure 5 compares the optimized tortuosities ( $\tau$ ) to obtain the best fit of the Guarracino and Jougnot (2018) model for each sample showed on Figs. 3 and 4 to the predicted tortuosities ( $\tau_e$ ) using Eq. (29). One can see that the best fit tortuosities fall very close to the 1:1 line showed here for reference, therefore indicating that Eq. (29) provides a fair approximation for the hydraulic tortuosity when it cannot be obtained otherwise.

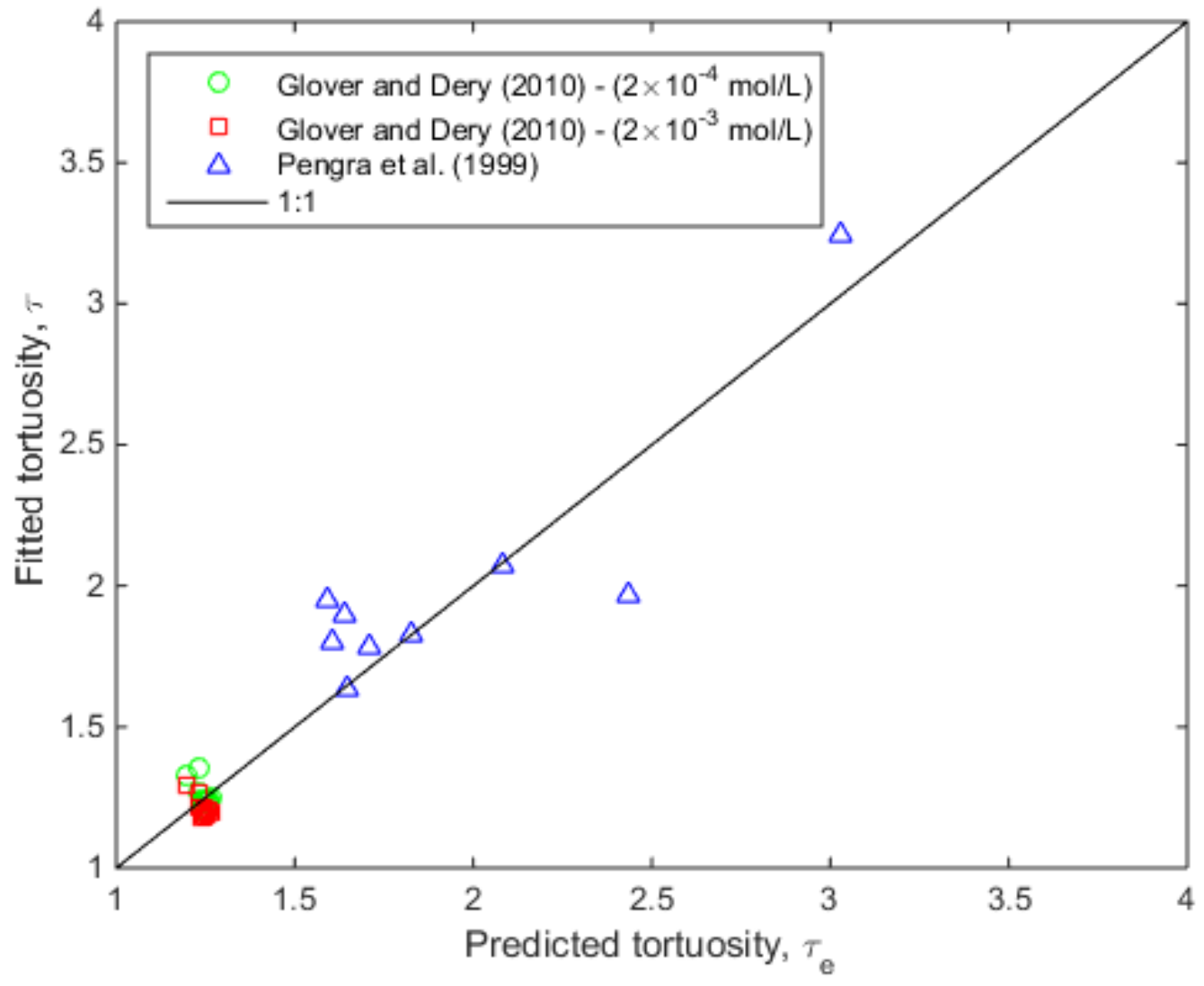

*Figure 5: Predicted versus best-fit tortuosities for the data from Glover and Déry (2010) and Pengra et al. (1999). The plain black line corresponds to 1:1 values (i.e., $\tau = \tau_e$).*

### 3.3 Evolution with saturation

The effect of the saturation on the effective excess charge density remains a vivid area of investigation as explained in the theory section. In the present chapter we compare the volume averaging approach of Linde et al. (2007) with the flux averaging approach of Jougnot et al. (2012). Figure 6a and b show the evolution of relative excess charge densities as a function of the effective saturation for the Jougnot et al. (2012) model (Eq. 26) using a pore size distribution inferred from the water retention ($f_D^{WR}$) and the relative permeability ($f_D^{RP}$) curves, respectively. The black lines correspond to the volume averaging approach for the corresponding soils. Note that the x-axis represents the effective saturation, defined as:

$$S_e = \frac{S_w - S_w^r}{1 - S_w^r}, \tag{30}$$

to remove the effect of the residual water saturation $S_w^r$ differences between the soil types. It explains why all the volume averaging curves are not superposed.

It can be noted that the effective excess charge always increases as the water saturation decreases. For the volume averaging model, it is due to decreasing volume of water in the pores while the amount of charges to compensate remains constant. For the flux averaging model, it is due to the fact that larger pores (i.e., smaller relative volume of EDL in the capillary) are desaturating first, letting the water flow through the smaller pores (i.e., smaller relative volume of EDL in the capillary). Hence, the model proposed by Jougnot et al. (2012) yields a soil-specific function $\hat{Q}_v^{\mathrm{REV}}(S_w)$ which strongly depends on the soil texture and shows very important changes with saturation, i.e., up to 9 orders of magnitudes (see also Soldi et al.. 2019).

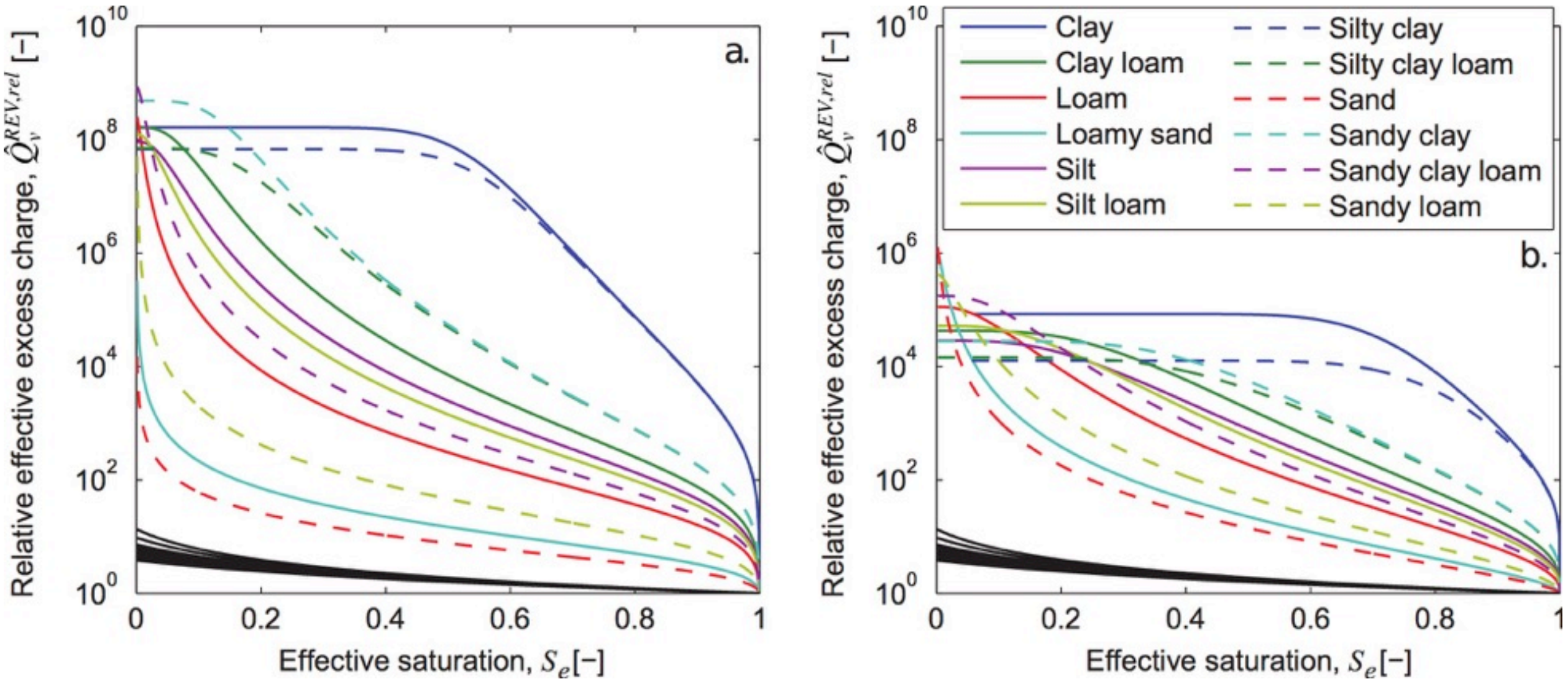

*Figure 6: Effective excess charge density of various soil types as a function of the saturation (modified from Jougnot et al. 2012).*

## 4. Pore network determination of the effective excess charge density

The present section describes a numerical up-scaling procedure to determine the effective excess charge density in a synthetic 2D pore network.

### *4.1 Equations of coupled fluxes in a single capillary*

Following the formalism exposed in Bernabé (1998), the hydraulic flux $Q$ and the electrical flux $J$ in a single capillary of radius $r$ and length $l$ are given by the two coupled equations:

$$\begin{cases} Q = -\dfrac{\pi r^4}{8\eta}\dfrac{(P_u - P_d)}{l} + \dfrac{\pi\varepsilon\varepsilon_0 r^2\zeta}{\eta}\left(1 - \dfrac{2}{r^2\zeta}\displaystyle\int_0^r r\psi(r)\,dr\right)\dfrac{(V_u - V_d)}{l} \\ J = \dfrac{\pi\varepsilon\varepsilon_0 r^2}{\eta}\left(1 - \dfrac{2}{r^2\zeta}\displaystyle\int_0^r r\psi(r)\,dr\right)\dfrac{(P_u - P_d)}{l} \\ \quad -\left[\dfrac{2\pi\varepsilon^2\varepsilon_0^2}{\eta}\displaystyle\int_0^r r\left(\dfrac{d\psi(r)}{dr}\right)^2 dr + 2\pi\sigma_f\displaystyle\int_0^r r\cosh\left(\dfrac{ze\psi(r)}{kT}\right)dr\right]\dfrac{(V_u - V_d)}{l} \end{cases} \quad (31)$$

where $P_u$ (resp. $V_u$) is the upstream hydraulic pressure (resp. the electrical potential) and $P_d$ (resp. $V_d$) the downstream pressure (resp. potential). The computation of the local electrical potential distribution $\psi$ inside the capillary is obtained by solving the Poisson-Boltzmann equation inside infinite cylinders, as done by Leroy and Maineult (2018).

The set of Eqs. (31) can be written as:

$$\begin{cases} Ql = -\gamma^h(P_u - P_d) + \gamma^c(V_u - V_d) \\ Jl = \gamma^c(P_u - P_d) - \gamma^e(V_u - V_d) \end{cases} \quad (32)$$

where $\gamma^h$ is the modified hydraulic conductance (in m$^4$ Pa$^{-1}$ s$^{-1}$), $\gamma^e$ the electrical conductance (in S m), and $\gamma^c$ the coupling conductance (in m$^4$ V$^{-1}$ s$^{-1}$).

### *4.2 2D tube network and linear system for the pressure and the electrical potential*

We consider a square random tube network as depicted in Fig. 7, for which all tubes are of length $l$ (in m).

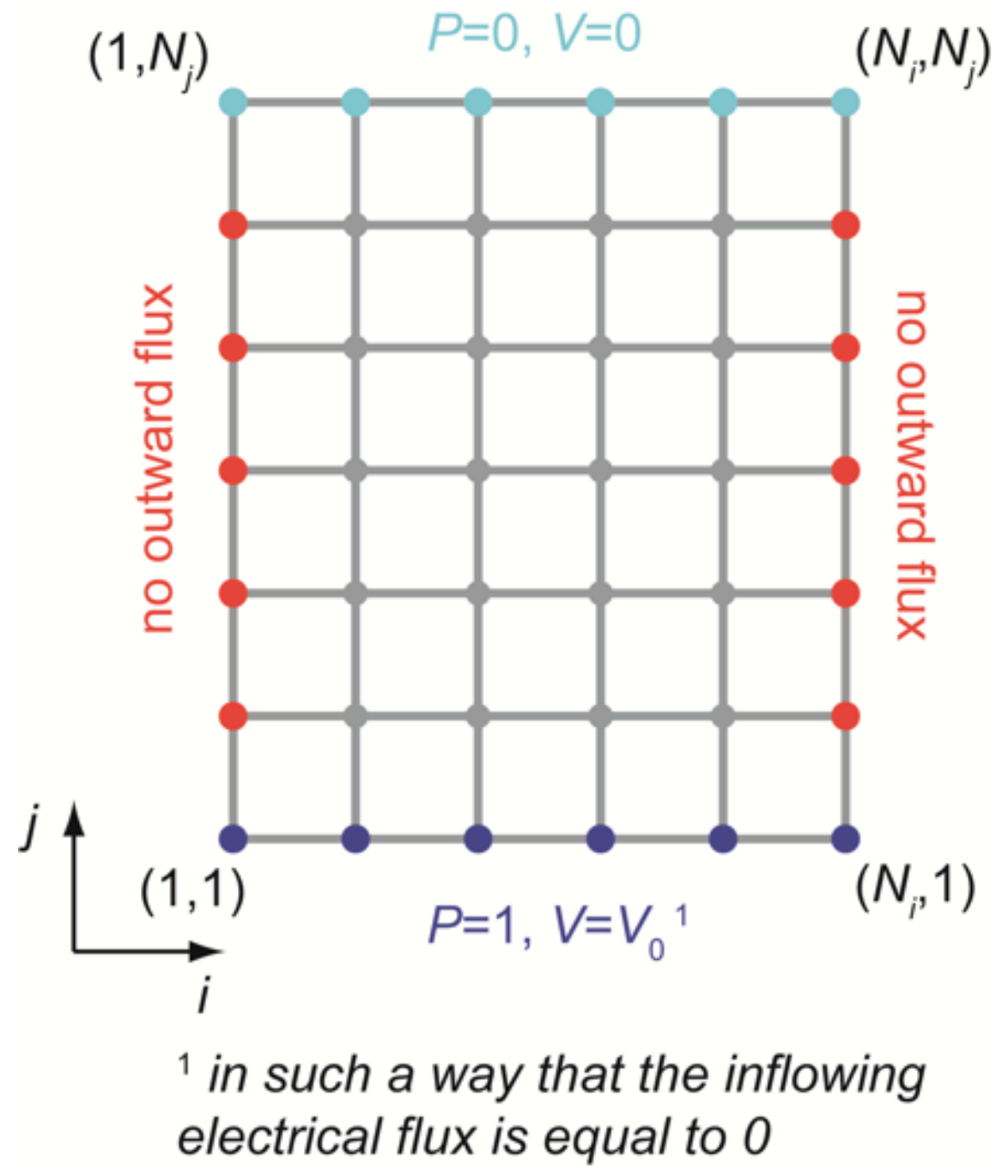

*Figure 7: Tube network and boundary conditions.*

Writing the conservation laws (Kirchhoff's laws, 1845) for the hydraulic flux and the electrical flux at each node of the network, combined with the appropriate boundary conditions, provides a linear system to be solved, whose unknown are the hydraulic pressures and electrical potential $P_{i,j}$ and $V_{i,j}$ at all nodes and the electrical potential $V_0$ (for more details, see Appendix A).

### *4.3 Computation of the petrophysical parameters*

The electrokinetic coupling coefficient (in V Pa$^{-1}$) is computed using:

$$C^{EK} = \frac{\Delta V}{\Delta P} = \frac{0 - V_0}{0 - 1} = V_0 . \tag{33}$$

The excess of charge density is given by reorganizing Eq. (18):

$$\hat{Q}_v = -\frac{\eta \sigma C^{EK}}{k} , \tag{34}$$

Neglecting the surface conductivity and introducing the formation factor gives:

$$\hat{Q}_v = -\frac{\eta \sigma_w C^{EK}}{kF} . \tag{35}$$

For the computation of the quantities $k\phi^{-1}$ and $F\phi$, see Appendix B.

*4.4 Applications*

We ran computations on uncorrelated random networks (i.e., the distribution of the tube radii is totally uncorrelated) of size 100 by 100 nodes (19800 tubes). We used a distribution such that the decimal logarithm of the radius is normally distributed, as done by Maineult et al. (2017) – see Fig. 8. The probability $P$ that log($r$) is less than $X$ is given by:

$$P\left(\log\left(r\right) \leq X\right) = \frac{1}{2} + \frac{1}{2}\,\mathrm{erf}\left(\frac{X - \log\left(r_{peak}\right)}{\mathrm{SD}\sqrt{2}}\right) \tag{36}$$

where SD is the standard deviation. We explored different values of $r_{peak}$ (i.e., 0.1, 0.2, 0.3, 0.5, 1, 2, 3, 5 and 10 μm), and took SD=0.5.

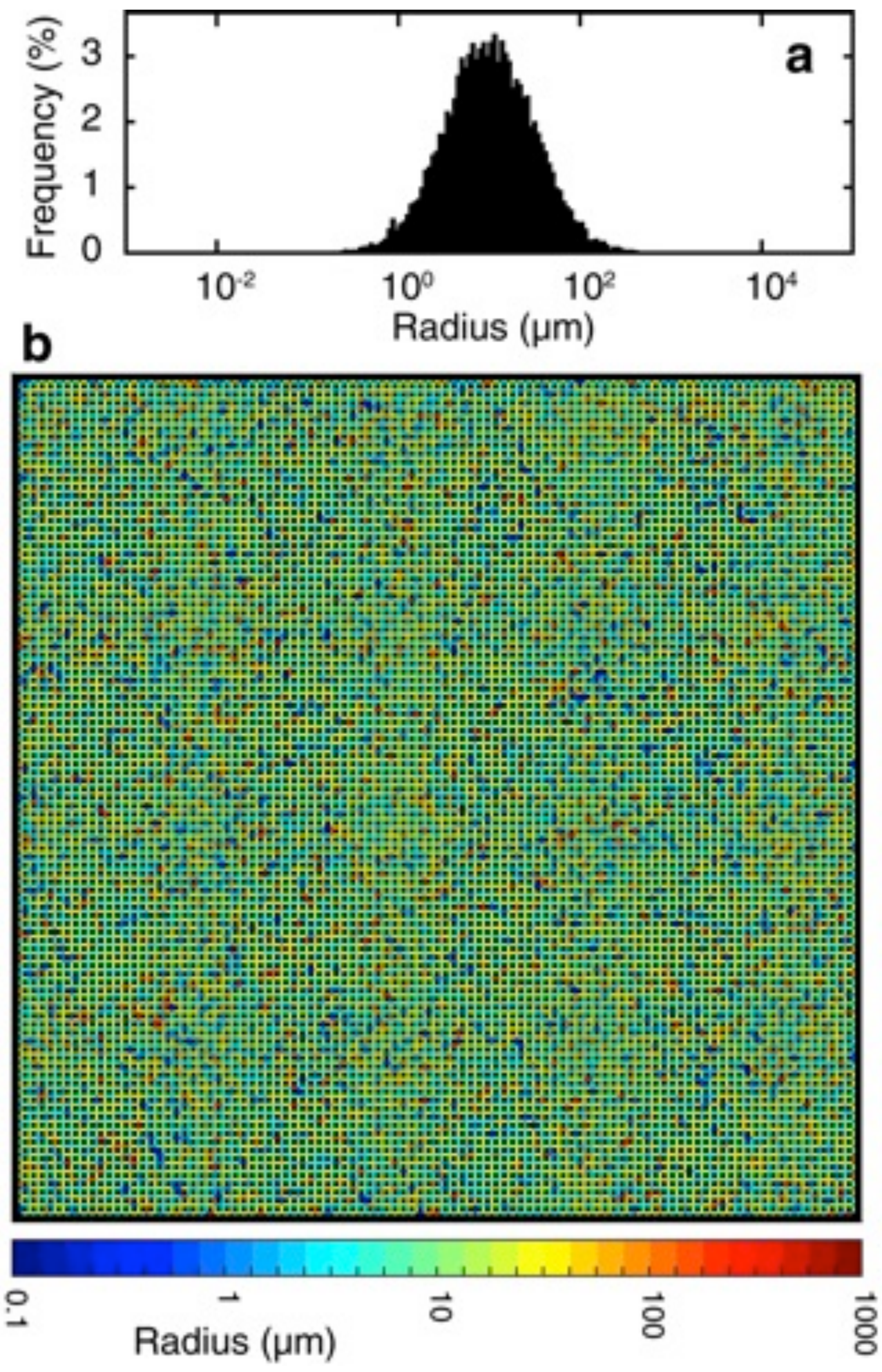

*Figure 8: Example of random uncorrelated media. Experimental distribution (a) of the tube radii (the decimal logarithm of the pore tube radius distribution is normally distributed, with a mean radius of 10 μm and a standard deviation of 0.5) associated with the network (100×100 nodes, 19800 tubes) shown in b (modified from Maineult et al., 2017).*

Note that to compute the fluid conductivity $\sigma_w$ associated with the concentration $C_w^0$, we used the empirical relation given by Sen and Goode (1992) for NaCl brine:

$$\sigma_w = \left(5.6 + 0.27T - 1.5\,10^{-4}T^2\right)M - \frac{2.36 + 0.099T}{1 + 0.214M}M^{\frac{3}{2}} \tag{37}$$

where $T$ is the normal (i.e., not absolute) temperature (in °C) and $M$ is the molality (in mol kg$^{-1}$). To convert the concentration $C_w^0$ into molality, we use the CRC Handbook Table at 20°C (Lide 2008). The $\zeta$-potential is then obtained from the relation given by Jaafar et al (2009) (Eq. 28).

Figure 9 shows the electrokinetic coupling coefficients calculated for different 2D pore networks having different permeabilities. For ionic concentrations larger than 0.01 mol/L, the coupling coefficient appears not to be dependent on the permeability despite the influence of the permeability in its definition (Eq. 18). This is a result of the linearly dependence on the permeability of the effective excess charge density, canceling the permeability in Eq. (18). That can be clearly seen in Fig. 10, where the analytical model of Guarracino and Jougnot (2018) predicts accurately the evolution of the effective excess charge density for the synthetic 2D pore network. Note that this very good fit is obtained from all the calculated parameters, with only one unknown, which has been fitted: $\tau = 2.3$.

Then, for 0.001 mol/L, the coupling coefficient tends to decrease for the lowest permeabilities (below $10^{-12}$ m$^2$), i.e., the smallest pore sizes, which also correspond to the poorer fit of Eq. (24) on the synthetic data. This can be expected from the assumptions of the Guarracino and Jougnot (2018) model which is only valid when the EDL thickness is small enough in comparison to the pore size (see discussion in Jougnot et al., 2019). Low permeabilities and low salinities therefore show a limitation of their model, as the local potential distribution in the EDL must be computed by solving the Poisson-Boltzmann equation (see Leroy and Maineult, 2018).

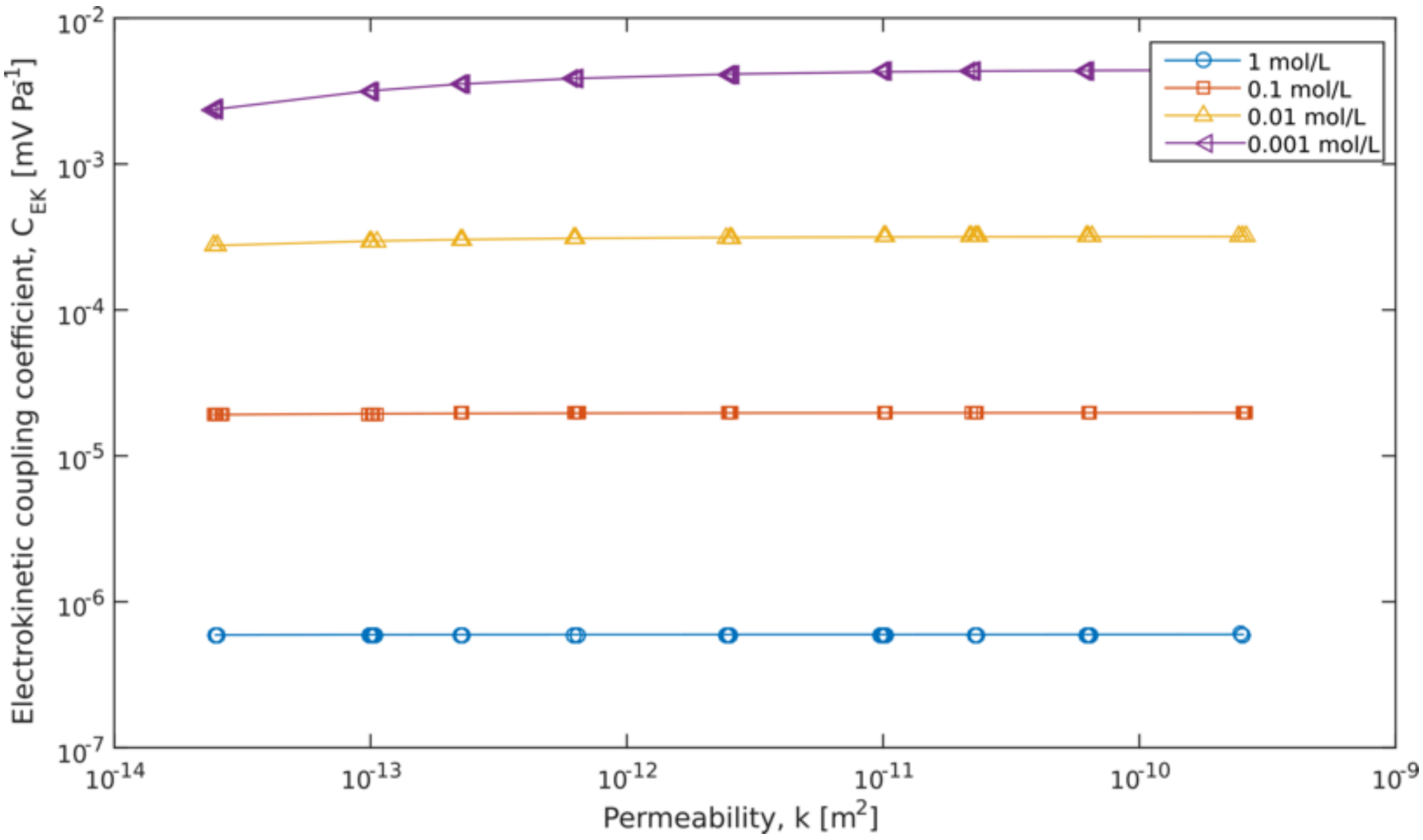

*Figure 9: Coupling coefficient of the 2D pore networks as a function of permeability for different NaCl concentrations.*

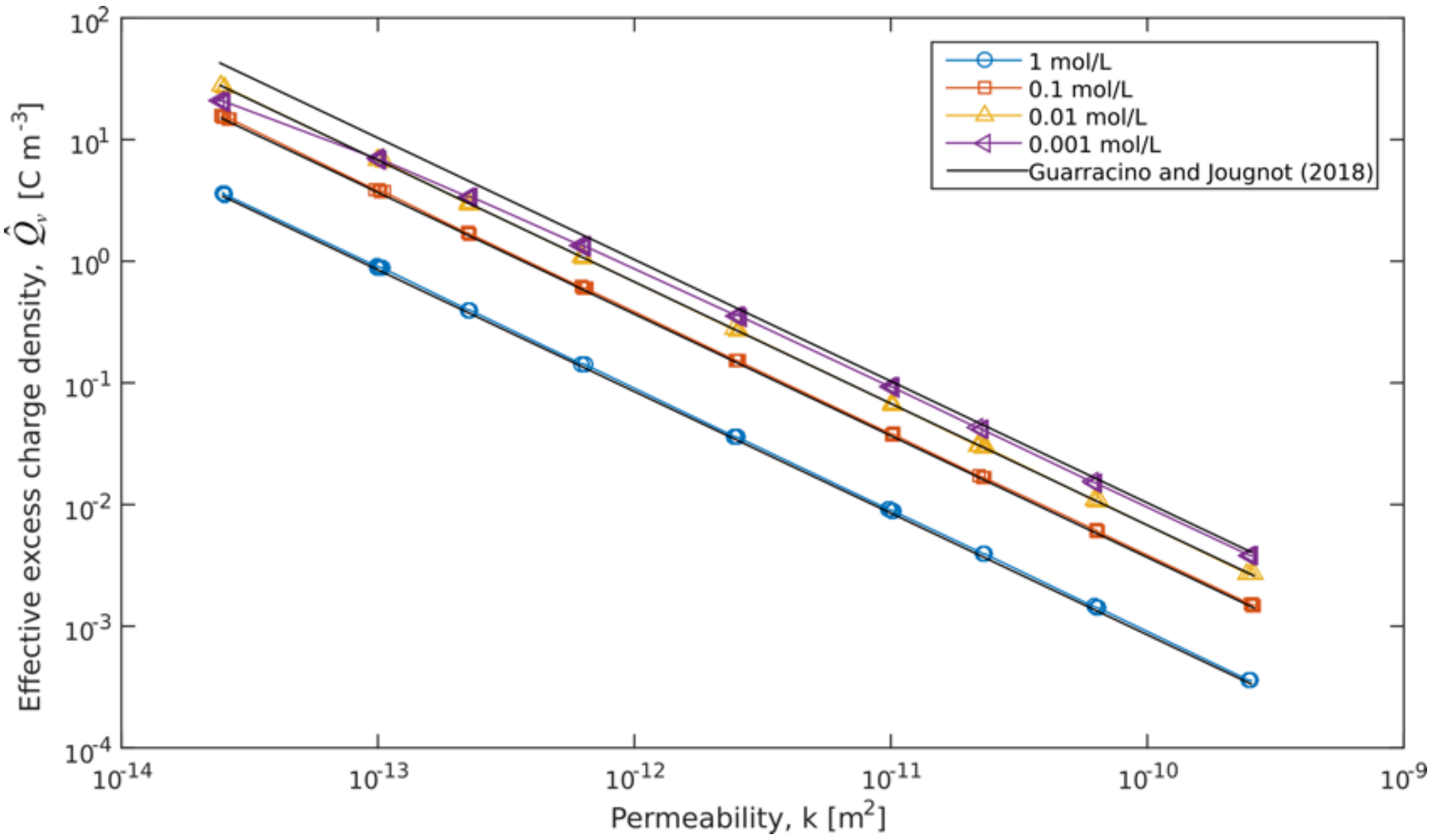

*Figure 10: Evolution of the excess charge density as a function of permeability for different NaCl concentrations: comparison between the 2D pore network results and model predictions of Guarracino and Jougnot (2018) for the corresponding ionic concentrations and* $\tau = 2.3$.

## 5. Use of the effective excess charge in numerical simulations

The present section illustrates the usefulness of the effective excess charge approach to model the streaming potential distribution in two kinds of complex media: a partially saturated soil and a fractured aquifer.

### *5.1 Rainwater infiltration monitoring*

Figure 11 describes the numerical framework that we use to simulate the streaming potential distribution resulting from a rainfall infiltration in a sandy loam soil. As explained in the theory section, the results of the hydrological simulation are used as input parameters for the electrical problem. In that scheme, it is clear that the electrokinetic coupling parameter is the effective excess charge density even if the water saturation distribution also plays a role through the electrical conductivity, affecting the amplitude of the SP signals.

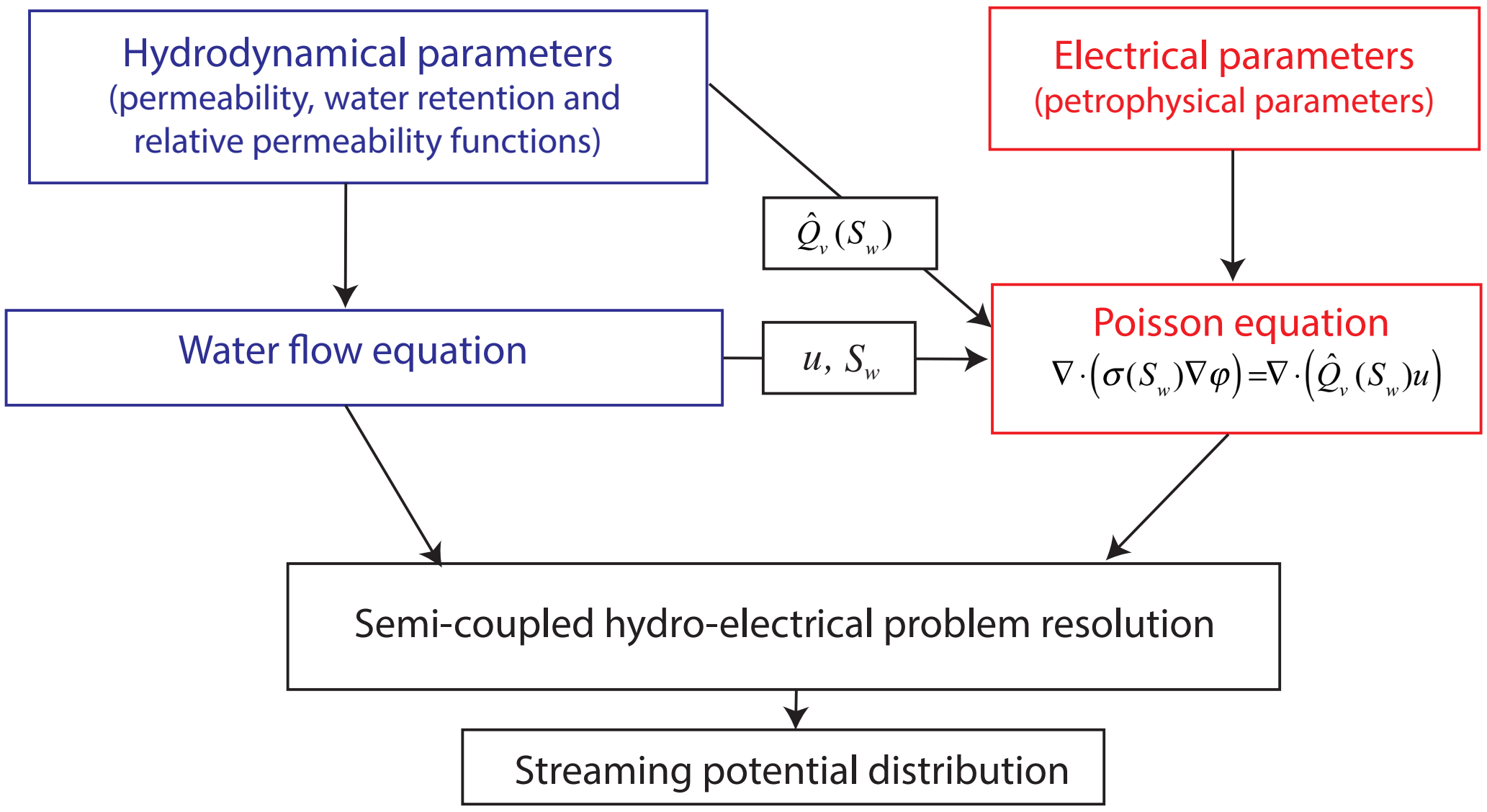

*Figure 11: Numerical framework for the simulation of the streaming potential distribution in a partially saturated porous medium.*

We consider a homogeneous sandy loam soil subjected to a rainfall event (Fig. 12a). The initial hydraulic conditions of the soil are set to hydrostatic equilibrium with a water table localized at 2.5 m depth. Following the work of Jougnot et al. (2015), the hydrological problem is solved using Hydrus 1D. This code solves the Richards equation to determine the evolution of the water saturation (Fig. 12b) and Darcy velocity as a function of depth and time. We choose the van Genuchten model to describe the water retention and the relative permeability functions, using the average hydrodynamic properties for a sandy loam soil proposed by Carsel and Parrish (1988).

The electrical problem is solved using a home-made code (for details please refer to Jougnot et al., 2015). As illustrated in Fig. 11, the hydrological simulation ouputs (i.e., the water saturation and the Darcy velocity distribution in both space and time) are used as input parameters for the electrical problem. The electrical conductivity is determined using Archie (1942) with the following petrophysical parameters: $m = 1.40$ the cementation exponent and $n = 1.57$ the saturation exponent. The effective excess charge is determined using Eq. (27) in which $\hat{Q}_v^{\mathrm{REV,rel}}(S_w)$ can be obtain from the WR or the RP flux averaging approach of Jougnot et al. 2012, or using the volume averaging approach of Linde et al. (2007) as explained in Section 2.4.2 (Fig. 13a).

Figure 12c shows the results of the numerical simulation of the streaming potential for virtual electrodes localized at different depths in the soil. Note that the reference electrode is localized at a depth of 3˚m. As the rainwater infiltration front progresses in the soil, the SP signals starts to increase. An electrode localized at the soil surface should be able to capture the highest signal amplitude during the rainfall, while the deeper electrodes show a time shift related to the time needed for the water flow to reach the electrode. The signal amplitude also decreases with depth as the Darcy velocity diminishes during the infiltration. The multimodal nature of the rainfall also vanishes, showing only a single SP peak at a depth of 5 cm. The $\hat{Q}_v(S_{\mathrm{w}})$ function used to plot Fig. 12c is the RP approach from Jougnot et al. (2012). Figure 13b shows the strong influence of the chosen approach on the vertical distribution of the signal amplitude at two different times ($t = 2$ and 10 d). These results are consistent with the findings of Linde et al. (2011), that is, the volume averaging model of Linde et al. (2007) does not allow to reproduce the large vertical SP signals that can be found in the literature (e.g., Doussan et al., 2002; Jougnot et al., 2015).

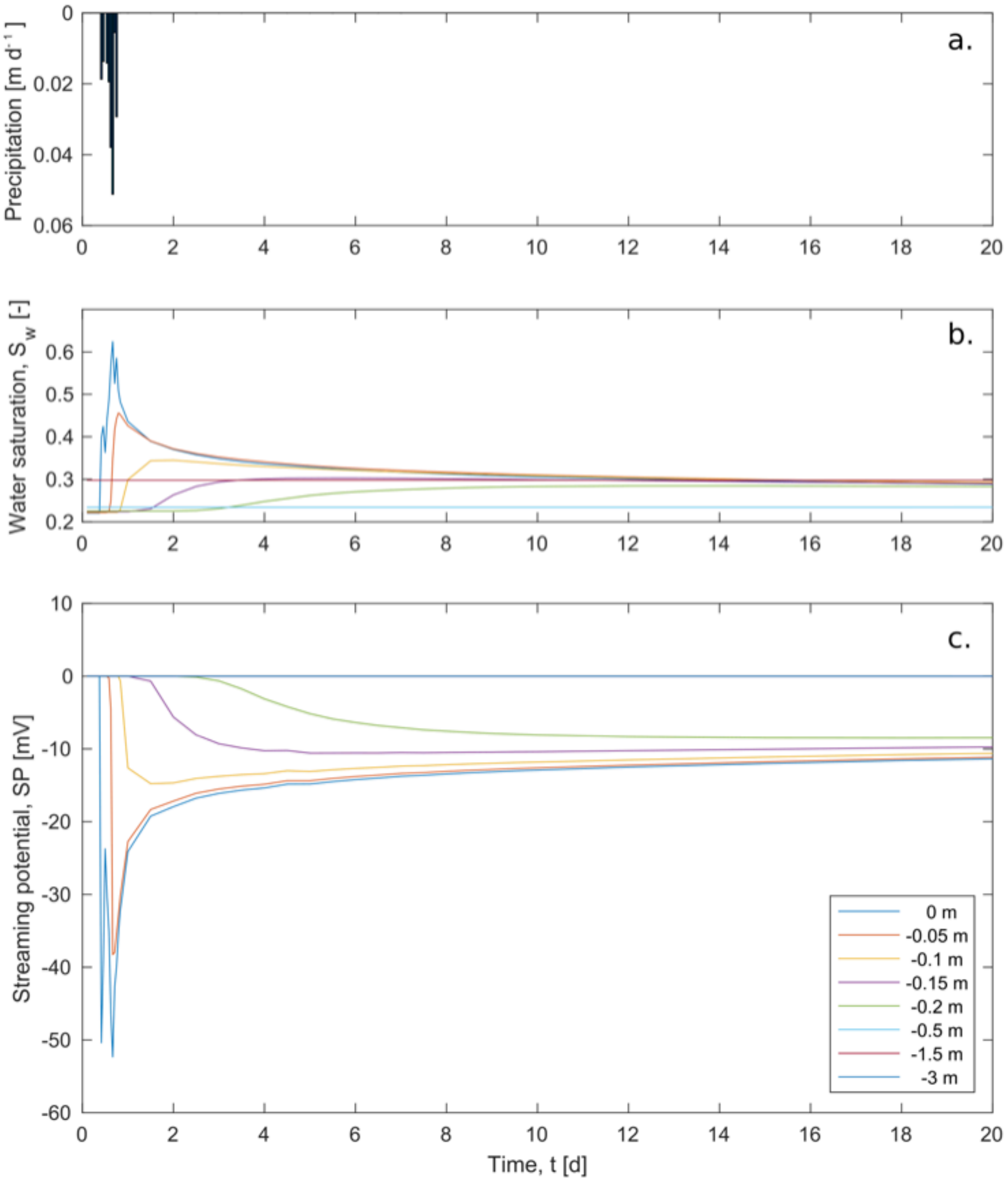

*Figure 12: Simulation results of the rainwater infiltration: (a) precipitation, (b) water saturation, and (c) streaming potential as a function of time.*

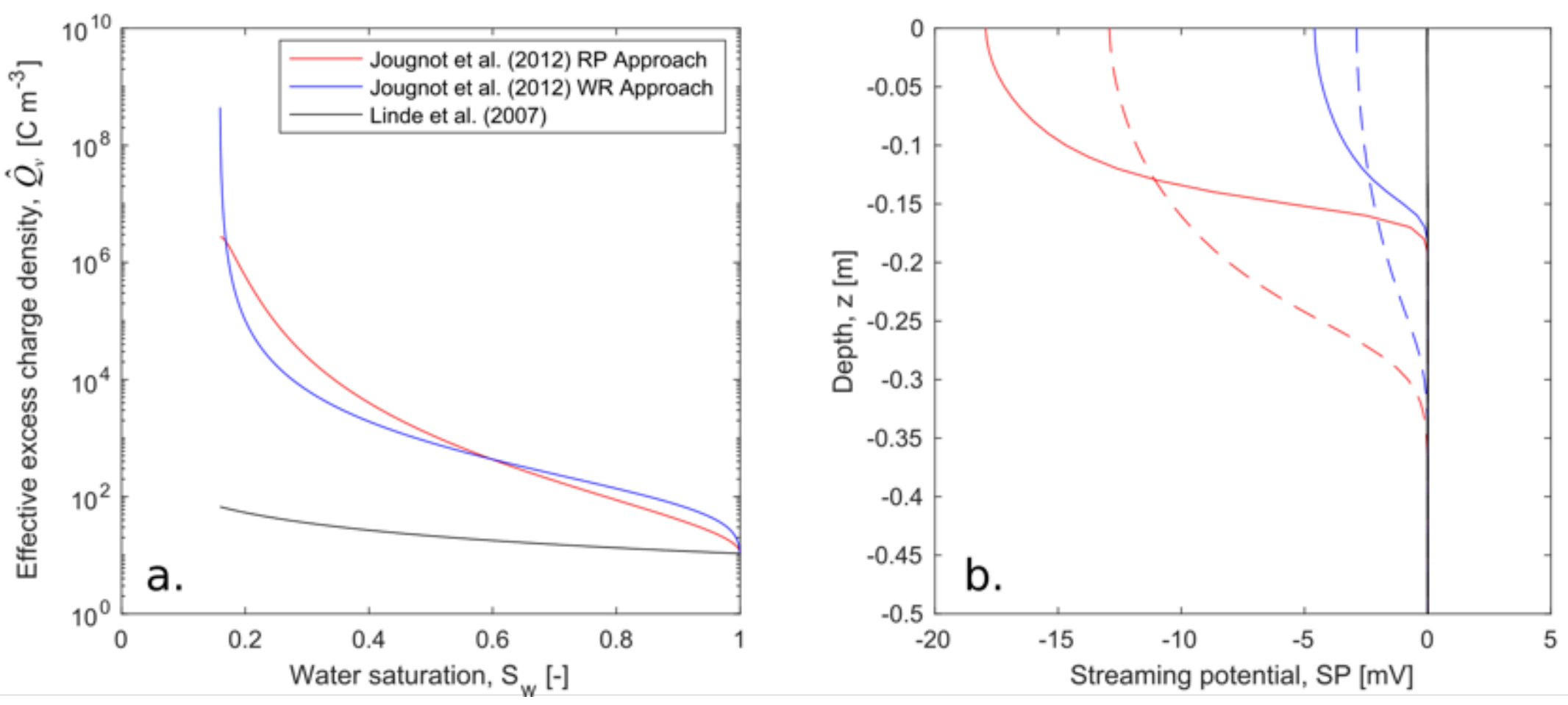

*Figure 13: (a) Comparison of the effective excess charge density as a function of the water saturation using Jougnot et al. (2012) RP and WR approaches and Linde et al. (2007). (b) Vertical distribution of the SP signal resulting for the rainwater infiltration using the corresponding $\hat{Q}_v(S_w)$ function at two different times, t =2 and 10 d, for the plain and the dashed lines, respectively.*

### *5.2 Pumping in a fractured medium*

The effective excess charge can also be used for modeling the streaming potential arising from groundwater flow in fractured media (e.g., Fagerlund and Heinson, 2003; Wishart et al., 2006; 2008; Maineult et al., 2013). Existing studies focusing on this phenomenon in fractured rocks suggest that monitoring the corresponding streaming potential under pumping conditions can help to identify the presence of fractures that interact with the surrounding matrix (Roubinet et al., 2016; DesRoches et al, 2017). This has been demonstrated with numerical approaches relying on a discrete representation of the considered fractures that are coupled to the matrix by using either the finite element method with adapted meshing (DesRoches et al, 2017) or the finite volume method within a dual-porosity framework (Roubinet et al., 2016).

The latter method is used here to illustrate the sensitivity of SP signals to hydraulically active fractures, and in particular to fractures having important fracture-matrix exchanges. For this purpose, we consider the coupled fluid flow and streaming potential problem described in Figure 11 that we apply to fractured porous domains under saturated conditions. In this case, the fluid flow problem is solved by considering Darcy's law and Darcy-scale mass conservation under steady-state conditions, and the effective excess charge is evaluated from the fracture and matrix permeability by adapting the strategy proposed in Jougnot et al. (2012) to two infinite plates having known separation and using the empirical relationship defined by Jardani et al. (2007), respectively. As shown in Roubinet et al. (2016), both fluid flow and streaming current must be simulated in the fractures and matrix to adequately solve this problem, even if the matrix is characterized by a very low permeability. Furthermore, relatively small fracture densities should be considered in order to individually detect the fractures that are hydraulically active.

Figure 14a, b, and c show three examples of fractal fracture network models defined by Watanabe and Takahashi (1995) for characterizing geothermal reservoirs and used in Gisladottir et al. (2016) for simulating heat transfer in these reservoirs. In these models, the number of fractures and the relative fracture lengths (i.e., the ratio of fracture to domain length) are defined from the fracture density, the smallest fracture length, and the fractal dimension that are set to 2.5, 0.1, and 1 m, respectively, considering a square domain of length $L = 100$ m. The positions of these fractures are randomly distributed, their angle can be equal to $\theta_1 = 25°$ or $\theta_2 = 145°$ with equal probability, and their aperture is set to $10^{-3}$ m. Note that we also add a deterministic fracture whose center is located at the domain center and whose angle is set to $\theta_1$ (represented in red in Figs. 14a-c). Finally, the fracture and matrix conductivity are set to $5\times10^{-2}$ and $5\times10^{-4}$ S m$^{-1}$, respectively, and the matrix permeability to $10^{-15}$ m$^2$.

The fluid flow and streaming potential problem is solved by considering (i) a pumping rate of $10^{-3}$ m$^3$ s$^{-1}$ applied at the domain center, (ii) gradient head boundary conditions with hydraulic head set to 1 and 0 m on the left and right sides of the domain, respectively, and (iii) a current insulation condition on all borders. Figure 1 shows the resulting difference of potential $\Delta\varphi_{x,y}$ and $\Delta\varphi_r$ where the white (Figs. 14d-f) and black (in Figs. 14g-i) dots represent the two largest SP signals measured along the dashed white circles that are plotted in Figs. 14d-f. These results show that a strong SP signal is observed for the primary fracture in which the pumping rate is applied when this fracture is not intersected by secondary fractures that are close to the pumping well (Figs. 14d and g). On the contrary, when the primary fracture is intersected by secondary fractures that are close to the pumping well and not connected to the domain borders, strong SP signals are observed at the extremities of the single secondary fracture (Figs. 14e and h) or the pair of secondary fractures (Figs. 14f and i). As demonstrated in existing studies (DesRoches et al, 2017; Roubinet et al., 2016), these results suggest that strong SP signals are associated with hydraulically active fractures, and that the largest values of SP measurements are related to important fracture-matrix exchanges.

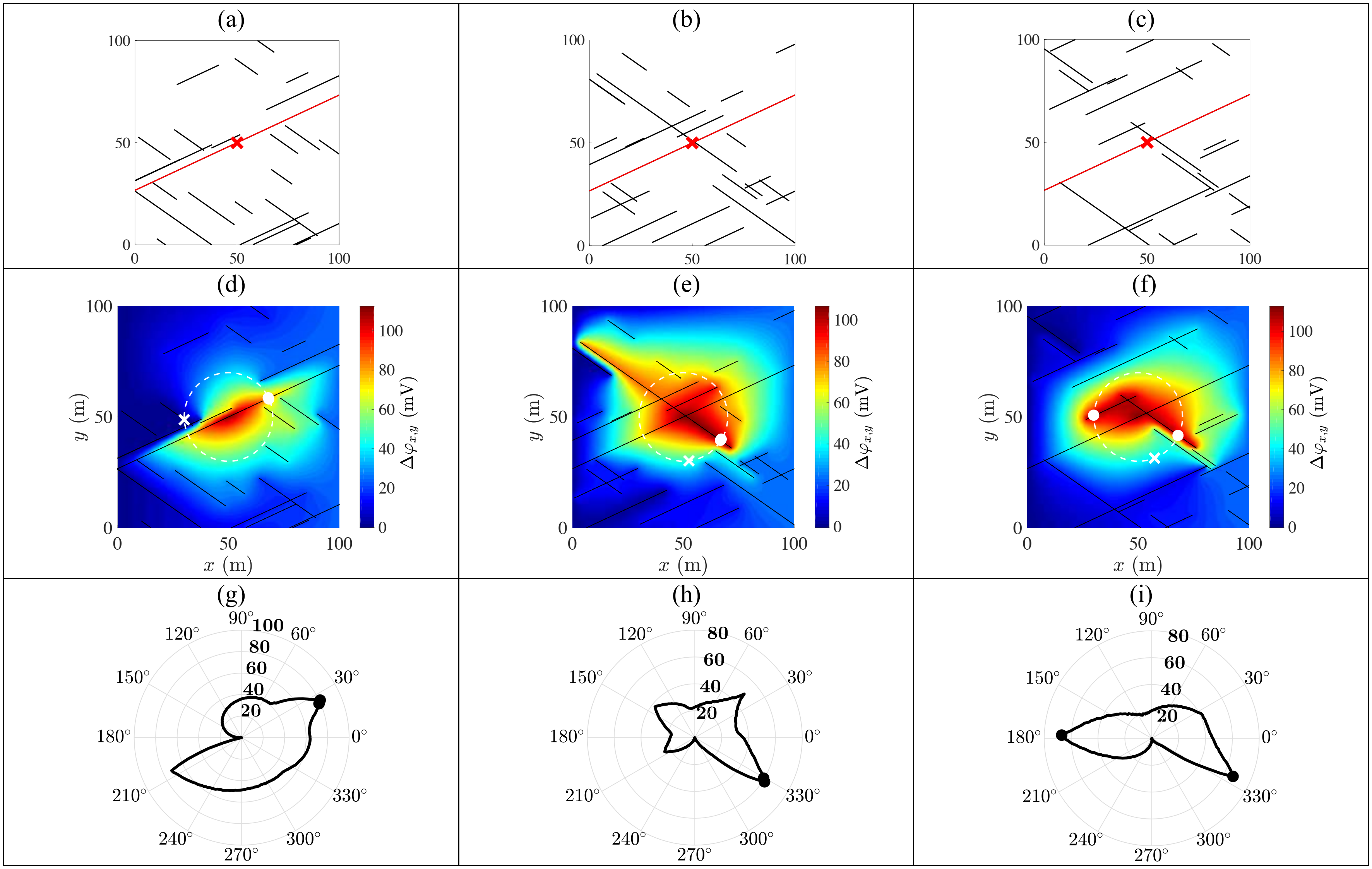

Figure 14 – (a-c) Studied fractured domains where the red cross represents the position of the considered pumping well. (d-f) Spatial distribution of the SP signal $\Delta\varphi_{x,y}$ (in mV) with respect to a reference electrode located at position $(x,y)=(0,0)$. (g-i) Polar plots of the SP signal $\Delta\varphi_r$ (in mV) along the dashed white circle plotted in (d-f) with respect to the minimum value measured along this circle and represented with a white cross.

## 6. Discussion and conclusion

Modeling of the streaming current generation and the corresponding electrical field requires a good understanding of electrokinetic coupling phenomena that occur when the water flows in porous and fractured media. This modeling can be done with two electrokinetic coupling parameters: coupling coefficient and effective excess charge. In this chapter we focused on the latter.

Considering the effective excess charge approach is quite recent (Kormiltsev et al. 1998) compare to the use of the coupling coefficient. Unlike the coupling coefficient, the effective excess charge density shows a strong dependence on petrophysical parameters (permeability, porosity, ionic concentration in the pore water). This has been highlighted by both empirical (Titov et al. 2001; Jardani et al. 2007) and mechanistic (Jougnot et al., 2012; Guarracino and Jougnot, 2018) approaches. The mechanistic approaches that we discuss in this chapter are based on the up-scaling process called flux-averaging as they propose an effective value for the excess charge density which is related to pore scale properties of the EDL and how the water flows through it.

Under saturated conditions, Guarracino and Jougnot (2018) model shows a linear dependence with geometrical properties (permeability, porosity, hydraulic tortuosity) and non-linears ones to chemical properties (ionic concentration, zeta potential). In section 3 and 4, we show that is provides good match with published laboratory data for various types of media as long as the model assumptions are respected (i.e., the pore radius should 5 times larger than the Debye length). The numerical simulations of 2D synthetic porous networks following the approaches of Bernabé (1998) and Maineult et al. (2017) strongly confirm these dependences.

Under partially saturated conditions, Jougnot et al. (2012) model shows a strong dependence of the effective excess charge density on the medium hydrodynamic properties of the porous medium. The function $\hat{Q}_v(S_w)$ becomes medium dependent and generally increases when the saturation decreases (up to 9 orders of magnitude).

The effective excess charge density approach has proven to be fairly useful to model the SP signal generation in complex media. In this chapter, we illustrate that with two examples: the SP monitoring of a rainfall infiltration and the SP response to pumping water in a fractured aquifer. In both cases the use of the effective excess charge as electrokinetic coupling parameter makes it simple to directly relate the streaming current generation to the water flux distribution in the medium and to explicitly take into account the medium heterogeneities below the REV scale (due to, for instance, saturation distribution, fractures). We believe that the development of that approach will help developing the use and modeling of streaming potentials in all kinds of media.

**Acknowledgment**

A. Maineult acknowledges the support of the Russian Science Foundation (Grant No. 17-17-01160 ‘Physical-chemical models of Induced Polarization and Self-Potential with application to pumping test experiments’).

## Appendix A: Equations for the pressure and electrical potential

This appendix details the calculation of the pressure and the electrical potential in the pore network. The Kirchhoff (1845) laws for the water flow and the electrical current at node of coordinates (*i*,*j*), which express the conservation of mass and the conservation of charge respectively, write:

$$
\begin{cases}
-\gamma^h_{i-1,j\to i,j}\left(P_{i,j}-P_{i-1,j}\right)+\gamma^c_{i-1,j\to i,j}\left(V_{i,j}-V_{i-1,j}\right)\\
-\gamma^h_{i+1,j\to i,j}\left(P_{i,j}-P_{i+1,j}\right)+\gamma^c_{i+1,j\to i,j}\left(V_{i,j}-V_{i+1,j}\right)\\
-\gamma^h_{i,j-1\to i,j}\left(P_{i,j}-P_{i,j-1}\right)+\gamma^c_{i,j-1\to i,j}\left(V_{i,j}-V_{i,j-1}\right)\\
-\gamma^h_{i,j+1\to i,j}\left(P_{i,j}-P_{i,j+1}\right)+\gamma^c_{i,j+1\to i,j}\left(V_{i,j}-V_{i,j+1}\right)=0\\
\gamma^c_{i-1,j\to i,j}\left(P_{i,j}-P_{i-1,j}\right)-\gamma^e_{i-1,j\to i,j}\left(V_{i,j}-V_{i-1,j}\right)\\
+\gamma^c_{i+1,j\to i,j}\left(P_{i,j}-P_{i+1,j}\right)-\gamma^e_{i+1,j\to i,j}\left(V_{i,j}-V_{i+1,j}\right)\\
+\gamma^c_{i,j-1\to i,j}\left(P_{i,j}-P_{i,j-1}\right)-\gamma^e_{i,j-1\to i,j}\left(V_{i,j}-V_{i,j-1}\right)\\
+\gamma^c_{i,j+1\to i,j}\left(P_{i,j}-P_{i,j+1}\right)-\gamma^e_{i,j+1\to i,j}\left(V_{i,j}-V_{i,j+1}\right)=0
\end{cases},
\tag{A1}
$$

for the node located in the interior of the network.

Inside the domain (i.e., for the indexes $(i,j) \in [2,N_i-1]\times[2,N_j-1]$), equations (A1) can be rewritten ;

$$
\begin{cases}
\gamma^h_{i-1,j\to i,j}P_{i-1,j}+\gamma^h_{i+1,j\to i,j}P_{i+1,j}-\kappa^h_{i,j}P_{i,j}+\gamma^h_{i,j-1\to i,j}P_{i,j-1}+\gamma^h_{i,j+1\to i,j}P_{i,j+1}\\
\quad -\gamma^c_{i-1,j\to i,j}V_{i-1,j}-\gamma^c_{i+1,j\to i,j}V_{i+1,j}+\kappa^c_{i,j}V_{i,j}-\gamma^c_{i,j-1\to i,j}V_{i,j-1}-\gamma^c_{i,j+1\to i,j}V_{i,j+1}=0\\
-\gamma^c_{i-1,j\to i,j}P_{i-1,j}-\gamma^c_{i+1,j\to i,j}P_{i+1,j}+\kappa^c_{i,j}P_{i,j}-\gamma^c_{i,j-1\to i,j}P_{i,j-1}-\gamma^c_{i,j+1\to i,j}P_{i,j+1}\\
\quad +\gamma^e_{i-1,j\to i,j}V_{i-1,j}+\gamma^e_{i+1,j\to i,j}V_{i+1,j}-\kappa^e_{i,j}V_{i,j}+\gamma^e_{i,j-1\to i,j}V_{i,j-1}+\gamma^e_{i,j+1\to i,j}V_{i,j+1}=0
\end{cases},
\tag{A2}
$$

with:

$$
\begin{cases}
\kappa^h_{i,j}=\left(\gamma^h_{i-1,j\to i,j}+\gamma^h_{i+1,j\to i,j}+\gamma^h_{i,j-1\to i,j}+\gamma^h_{i,j+1\to i,j}\right)\\
\kappa^c_{i,j}=\left(\gamma^c_{i-1,j\to i,j}+\gamma^c_{i+1,j\to i,j}+\gamma^c_{i,j-1\to i,j}+\gamma^c_{i,j+1\to i,j}\right),\\
\kappa^e_{i,j}=\left(\gamma^e_{i-1,j\to i,j}+\gamma^e_{i+1,j\to i,j}+\gamma^e_{i,j-1\to i,j}+\gamma^e_{i,j+1\to i,j}\right)
\end{cases}
\tag{A3}
$$

In $i=1$ (no outward flux), $j\in[2,N_j-1]$, we have (see Figure 1):

$$\begin{cases} \gamma^h_{2,j\to1,j}P_{2,j} - \kappa^h_{1,j}P_{1,j} + \gamma^h_{1,j-1\to1,j}P_{1,j-1} + \gamma^h_{1,j+1\to1,j}P_{1,j+1} \\ \qquad -\gamma^c_{2,j\to1,j}V_{2,j} + \kappa^c_{1,j}V_{1,j} - \gamma^c_{1,j-1\to1,j}V_{1,j-1} - \gamma^c_{1,j+1\to1,j}V_{1,j+1} = 0 \\ -\gamma^c_{2,j\to1,j}P_{2,j} + \kappa^c_{1,j}P_{1,j} - \gamma^c_{1,j-1\to1,j}P_{1,j-1} - \gamma^c_{1,j+1\to1,j}P_{1,j+1} \\ \qquad +\gamma^e_{2,j\to1,j}V_{2,j} - \kappa^e_{1,j}V_{1,j} + \gamma^e_{1,j-1\to1,j}V_{1,j-1} + \gamma^e_{1,j+1\to1,j}V_{1,j+1} = 0 \end{cases}, \tag{A4}$$

with:

$$\begin{cases} \kappa^h_{1,j} = \left(\gamma^h_{2,j\to1,j} + \gamma^h_{1,j-1\to1,j} + \gamma^h_{1,j+1\to1,j}\right) \\ \kappa^c_{1,j} = \left(\gamma^c_{2,j\to1,j} + \gamma^c_{1,j-1\to1,j} + \gamma^c_{1,j+1\to1,j}\right) \\ \kappa^e_{1,j} = \left(\gamma^e_{2,j\to1,j} + \gamma^e_{1,j-1\to1,j} + \gamma^e_{1,j+1\to1,j}\right) \end{cases}. \tag{A5}$$

In $i=N_i$ (no outward flux), $j\in[2,N_j-1]$ (see Figure 7), we have:

$$\begin{cases} \gamma^h_{N_i-1,j\to N_i,j}P_{N_i-1,j} - \kappa^h_{N_i,j}P_{N_i,j} + \gamma^h_{N_i,j-1\to N_i,j}P_{N_i,j-1} + \gamma^h_{N_i,j+1\to N_i,j}P_{N_i,j+1} \\ \qquad -\gamma^c_{N_i-1,j\to N_i,j}V_{N_i-1,j} + \kappa^c_{N_i,j}V_{N_i,j} - \gamma^c_{N_i,j-1\to N_i,j}V_{N_i,j-1} - \gamma^c_{N_i,j+1\to N_i,j}V_{N_i,j+1} = 0 \\ -\gamma^c_{N_i-1,j\to N_i,j}P_{N_i-1,j} + \kappa^c_{N_i,j}P_{N_i,j} - \gamma^c_{N_i,j-1\to N_i,j}P_{N_i,j-1} - \gamma^c_{N_i,j+1\to N_i,j}P_{N_i,j+1} \\ \qquad +\gamma^e_{N_i-1,j\to N_i,j}V_{N_i-1,j} - \kappa^e_{N_i,j}V_{N_i,j} + \gamma^e_{N_i,j-1\to N_i,j}V_{N_i,j-1} + \gamma^e_{N_i,j+1\to N_i,j}V_{N_i,j+1} = 0 \end{cases}, \tag{A6}$$

with:

$$\begin{cases} \kappa^h_{N_i,j} = \left(\gamma^h_{N_i-1,j\to N_i,j} + \gamma^h_{N_i,j-1\to N_i,j} + \gamma^h_{N_i,j+1\to N_i,j}\right) \\ \kappa^c_{N_i,j} = \left(\gamma^c_{N_i-1,j\to N_i,j} + \gamma^c_{N_i,j-1\to N_i,j} + \gamma^c_{N_i,j+1\to N_i,j}\right) \\ \kappa^e_{N_i,j} = \left(\gamma^e_{N_i-1,j\to N_i,j} + \gamma^e_{N_i,j-1\to N_i,j} + \gamma^e_{N_i,j+1\to N_i,j}\right) \end{cases}. \tag{A7}$$

In $j = 1$, the imposed hydraulic pressure is equal to 1:

$$P_{i,1} = 1, \tag{A8}$$

and the electrical potential is equal to $V_0$:

$$V_{i,1} = V_0, \tag{A9}$$

in such a way that the total entering electrical flux is equal to 0, that is to say:

$$\sum_{i=1}^{N_i} J_{i,1\to i,2}l = \sum_{i=1}^{N_i}\left(\gamma^c_{i,1\to i,2}\left(P_{i,2}-P_{i,1}\right) - \gamma^e_{i,1\to i,2}\left(V_{i,2}-V_{i,1}\right)\right) = 0, \tag{A10}$$

from which the following relation can be deduced:

$$\left(\sum_{i=1}^{N_i}\gamma^e_{i,1\to i,2}\right)V_0-\sum_{i=1}^{N_i}\gamma^c_{i,1\to i,2}P_{i,1}+\sum_{i=1}^{N_i}\gamma^c_{i,1\to i,2}P_{i,2}-\sum_{i=1}^{N_i}\gamma^e_{i,1\to i,2}V_{i,2}=0. \quad \text{(A11)}$$

Finally, in $j = N_j$, the imposed hydraulic pressure is equal to 0:

$$P_{i,N_j}=0, \quad \text{(A12)}$$

and the electrical potential is equal to 0 (potential gauge):

$$V_{i,N_j}=0. \quad \text{(A13)}$$

This set of equations forms a linear system, whose unknowns are the hydraulic pressures and electrical potential $P_{i,j}$ and $V_{i,j}$ at all nodes and the electrical potential $V_0$.

**Appendix B. Computation of the normal permeability and formation factor.**

In the classical case, the hydraulic flux $F_{x\to y}$ through a tube linking two nodes $x$ and $y$ writes (Poiseuille law):

$$F_{x\to y}=\frac{\pi r_{x\to y}^{\;4}}{8\eta}\frac{P_x-P_y}{l}=g^h{}_{x\to y}\left(P_x-P_y\right) \quad . \quad \text{(B1)}$$

We introduce the modified hydraulic flux $\Phi^h{}_{x\to y}$, eliminating the length $l$:

$$\Phi^h{}_{x\to y}=F_{x\to y}l=\frac{\pi r_{x\to y}^{\;4}}{8\eta}\left(P_x-P_y\right)=\gamma^h{}_{x\to y}\left(P_x-P_y\right) \quad . \quad \text{(B2)}$$

Under the assumption that the surface conductivity can be neglected, the electrical flux $J_{x\to y}$ writes:

$$J_{x\to y}=\sigma_f\pi r_{x\to y}^{\;2}\frac{V_x-V_y}{l}=g^e{}_{x\to y}\left(V_x-V_y\right) \quad . \quad \text{(B3)}$$

Them, we use the modified electrical flux $\Phi^e{}_{x\to y}$ to simplify from $l$ and the fluid conductivity $\sigma_f$

$$\Phi^e{}_{x\to y}=J_{x\to y}\frac{l}{\sigma_f}=\pi r_{x\to y}^{\;2}\left(V_x-V_y\right)=\gamma^e{}_{x\to y}\left(V_x-V_y\right) \quad \text{(B4)}$$

Kirchhoff's law (1845) applies at any node inside the square network (Figure 7):

$$Z_{i,j-1\to i,j} + Z_{i-1,j\to i,j} + Z_{i+1,j\to i,j} + Z_{i,j+1\to i,j} = 0 \tag{B5}$$

with $Z$ equal to $F$ or $J$ respectively. Using equation (B1) or (B3), this leads to:

$$\begin{aligned} &a_{i,j-1\to i,j}X_{i,j-1} + a_{i-1,j\to i,j}X_{i-1,j} - \left(a_{i,j-1\to i,j} + a_{i-1,j\to i,j} + a_{i+1,j\to i,j} + a_{i,j+1\to i,j}\right)X_{i,j} \\ &\quad + a_{i+1,j\to i,j}X_{i+1,j} + a_{i,j+1\to i,j}X_{i,j+1} = 0 \end{aligned} \tag{B6}$$

with $a = r^4$ and $X = P$ for the hydraulic case, and $a = r^2$ et $X = V$ for the electrical case.

For the nodes on the borders of the network, Eq. (B6) is easily modified to take into account the boundary conditions (i.e., no outward flow for $i = 1$ and $i = N_i$, $P = 1$ or $V = 1$ for $j = 1$, and $P = 0$ and $V = 0$ for $j = N_j$).

A linear system is obtained; the hydraulic pressure or electrical potential at the nodes of the network are the $N_iN_j$ unknowns. The modified fluxes can be computed after the system is solved through the use of equations (B2) and (B4).

One can the compute the permeability of the network using Darcy's law:

$$k = \frac{\eta Q L}{S\left|\Delta P\right|} = \frac{\eta}{l^2}\frac{N_j - 1}{N_i - 1}\frac{\Phi^h{}_{\Sigma out/in}}{\left|\Delta P\right|} \tag{B7}$$

where $Q$ is the hydraulic flux, $L$ is the length of the network along the flow direction (i.e., $j$-direction), $S$ the transversal section, and the total out-flowing and in-flowing fluxes are given by:

$$\begin{cases} \Phi^h{}_{\Sigma out} = \sum_{i=1}^{N_i-1} \Phi^h{}_{i,N_j-1\to i,N_j} \\ \Phi^h{}_{\Sigma in} = \sum_{i=1}^{N_i-1} \Phi^h{}_{i,1\to i,2} \end{cases}. \tag{B8}$$

To estimate $S$ and the porosity of the network, we extend the 2D network into a virtual 3D one, by adding at each node two vertical tubes of length $l/2$, which do not contribute to the transport. The section $S$ is thus equal to $(N_i–1)l^2$, and the porosity $\phi$ can be expressed as:

$$\phi = \frac{\left((N_i - 1)N_j + (N_i - 1)N_j + N_iN_j\right)\pi\left\langle r^2\right\rangle l}{(N_i - 1)(N_j - 1)l^3}. \tag{B9}$$

Extracting $l^{-2}$ from (B9) and reminding that $|\Delta P| = 1$, it comes:

$$\frac{k}{\phi} = \frac{\eta}{\pi\langle r^2\rangle} \frac{\left(N_j - 1\right)^2}{\left(N_i - 1\right)N_j + \left(N_i - 1\right)N_j + N_i N_j} \Phi^h{}_{\Sigma out/in}. \tag{B10}$$

The formation factor of the network is obtained through:

$$\frac{1}{F} = \frac{\sigma_r}{\sigma_w} = \frac{1}{\sigma_w}\frac{JL}{S|\Delta V|} = \frac{1}{l^2}\frac{N_j - 1}{N_i - 1}\frac{\Phi^e{}_{\Sigma out/in}}{|\Delta V|}. \tag{B11}$$

Taking into account that $|\Delta V| = 1$ it becomes:

$$\frac{1}{F\phi} = \frac{1}{\pi\langle r^2\rangle} \frac{\left(N_j - 1\right)^2}{\left(N_i - 1\right)N_j + \left(N_i - 1\right)N_j + N_i N_j} \Phi^e{}_{\Sigma out/in}. \tag{B12}$$